\documentclass[article]{IEEEtran} \newcommand{\figsize}{3.3}

\usepackage{graphicx,psfrag}
\usepackage[cmex10]{amsmath}
\usepackage{amsfonts,amssymb,latexsym,hyperref}
\usepackage{algorithm,algorithmic}

\usepackage{cite}
\usepackage[usenames,dvipsnames]{color}
\usepackage[normalem]{ulem} 

\usepackage{etoolbox}
\preto\subequations{\ifhmode\unskip\fi}

 \newcommand{\putFrag}[4]{\begin{figure}[t]
                            \centering
                            #4
                            \includegraphics[width=#3in]{figures/#1.eps}
                            \caption{#2}
                            \label{fig:#1}
                            \vspace{1mm}
                          \end{figure} }
 \newcommand{\putTable}[3]{\begin{table}[t]
                            \centering
                            #3
                            \vspace{2mm}
                            \caption{#2}
                            \vspace{-4mm}
                            \label{tab:#1}
                          \end{table} }
 \newcommand{\capFrag}[2]{}
 \newcommand{\capTable}[2]{}

 \renewcommand{\tilde}{\widetilde}
 \renewcommand{\hat}{\widehat}
 \renewcommand{\bar}{\overline}
 \newcommand{\defn}{\triangleq}

 \newcommand{\tvec}[1]{\ensuremath{\Tilde{\boldsymbol{#1}}}}
 \newcommand{\ovec}[1]{\ensuremath{\Bar{\boldsymbol{#1}}}}
 \newcommand{\hvec}[1]{\ensuremath{\Hat{\boldsymbol{#1}}}}
 
 \renewcommand{\vec}[1]{\ensuremath{\boldsymbol{#1}}}

 \newcommand{\norm}[1]{\ensuremath{\| #1 \|}}
 \newcommand{\mc}[1]{\ensuremath{\mathcal{#1}}}

 \newcommand{\Real}{{\mathbb{R}}}
 \newcommand{\Complex}{{\mathbb{C}}}

 \newcommand{\of}[1]{^{(#1)}}

 \newcommand{\tran}{^{\text{\textsf{T}}}}
 \newcommand{\herm}{^{\text{\textsf{H}}}}
 
 \newcommand{\true}{^0}
 \renewcommand{\j}{\mathrm{j}}

 \DeclareMathOperator{\sgn}{sgn}
 
 \DeclareMathOperator{\E}{E}
 \DeclareMathOperator{\var}{var}

 \DeclareMathOperator{\rank}{rank}

 \DeclareMathOperator{\Diag}{Diag}

 \renewcommand{\eqref}[1]{(\ref{eq:#1})}
 
 \newcommand{\Figref}[1]{Figure~\ref{fig:#1}}
 \newcommand{\figref}[1]{Fig.~\ref{fig:#1}}
 \newcommand{\tabref}[1]{Table~\ref{tab:#1}}
 \newcommand{\secref}[1]{Section~\ref{sec:#1}}
 
 \newcommand{\appref}[1]{Appendix~\ref{app:#1}}

 \newcommand{\algref}[1]{Algorithm~\ref{alg:#1}}
 \newcommand{\lineref}[1]{line~\ref{line:#1}}

 \newcounter{comment}[section]
 
 \newcounter{texthead}[section]

 \newcommand{\shrinkage}[1]{\vec{\eta}_{\sf #1\!}}
 \newcommand{\tied}{^{\text{\sf tied}}}
 \newcommand{\untied}{^{\text{\sf untied}}}
 \newcommand{\bkt}[1]{\langle #1 \rangle}
 \newcommand{\dvg}{\nu} 
 
 \newcommand{\Nr}{N_r}
 \newcommand{\Nu}{N_u}
 \newcommand{\Nc}{N_c}

\begin{document}
\setlength{\arraycolsep}{0.5mm}

\title{AMP-Inspired Deep Networks for Sparse Linear Inverse Problems}

\author{Mark Borgerding, Philip Schniter, and Sundeep Rangan%
\thanks{M.~Borgerding (email: borgerding.7@osu.edu) 
        and P.~Schniter (email: schniter.1@osu.edu)
        are with the Department of Electrical and Computer Engineering,
        The Ohio State University, Columbus OH.
        Their work was supported in part by 
        the National Science Foundation under grants 1527162 and 1539960.
        S.~Rangan (email: srangan@nyu.edu) 
        is with the Department of Electrical and Computer Engineering,
        New York University, Brooklyn, NY, 11201.
        His work was supported by the National Science
        Foundation under Grants 1302336, 1547332, and 1564142.}
\thanks{Portions of this work were presented at the 2016
        IEEE Global Conference on Signal and Information Processing
        \cite{Borgerding:GSIP:16}.}
        }

\maketitle

\begin{abstract}
Deep learning has gained great popularity due to its widespread success on many inference problems.
We consider the application of deep learning to the sparse linear inverse problem, where one seeks to recover a sparse signal from a few noisy linear measurements.
In this paper, we propose two novel neural-network architectures that decouple prediction errors across layers in the same way that the approximate message passing (AMP) algorithms decouple them across iterations: through Onsager correction.
First, we propose a ``learned AMP'' network that significantly improves upon Gregor and LeCun's ``learned ISTA.''
Second, inspired by the recently proposed ``vector AMP'' (VAMP) algorithm,
we propose a ``learned VAMP'' network that offers increased robustness to deviations in the measurement matrix from i.i.d.\ Gaussian.
In both cases, we jointly learn the linear transforms and scalar nonlinearities of the network.
Interestingly, with i.i.d.\ signals, the linear transforms and scalar nonlinearities prescribed by the VAMP algorithm coincide with the values learned through back-propagation, leading to an intuitive interpretation of learned VAMP. 
Finally, we apply our methods to two problems from 5G wireless communications: compressive random access and massive-MIMO channel estimation.
\end{abstract}

\begin{IEEEkeywords}
Deep learning, 
compressive sensing, 
approximate message passing, 
random access, massive MIMO.
\end{IEEEkeywords}

%

\section{Introduction} \label{sec:intro}

We consider the problem of recovering a signal $\vec{s}\true\in\Real^N$ from a noisy linear measurement $\vec{y}\in\Real^M$ of the form\footnote{Although we focus on real-valued quantities for ease of illustration, the methods in this paper could be easily extended to the complex-valued case.}
\begin{align}
\vec{y} = \vec{\Phi s}\true + \vec{w} 
\label{eq:ys} ,
\end{align}
where $\vec{\Phi}\in\Real^{M\times N}$ represents a linear operator and $\vec{w}\in\Real^M$ is additive white Gaussian noise (AWGN).
In many cases of interest, $M\ll N$.
We will assume that the signal vector $\vec{s}\true$ has an (approximately) sparse\footnote{Although we focus on sparse signals, the methods in this paper can be applied to other signals, such as the finite-alphabet signals used in digital communications.} representation in a known orthonormal basis $\vec{\Psi}\in\Real^{N\times N}$, i.e., that $\vec{s}\true=\vec{\Psi x}\true$ for some (approximately) sparse vector $\vec{x}\true\in\Real^N$.
Thus we define $\vec{A}\defn\vec{\Phi \Psi}\in\Real^{M\times N}$, write \eqref{ys} as
\begin{align}
\vec{y} = \vec{Ax}\true + \vec{w}
\label{eq:y} ,
\end{align}
and seek to recover a sparse $\vec{x}\true$ from $\vec{y}$.
In the sequel, we will refer to this problem as the ``sparse linear inverse'' problem.
The resulting estimate $\hvec{x}$ of $\vec{x}\true$ can then be converted into an estimate $\hvec{s}$ of $\vec{s}\true$ via $\hvec{s}=\vec{\Psi}\hvec{x}$.

The sparse linear inverse problem has received enormous attention over the last few years, in large part because it is central to compressive sensing \cite{Eldar:Book:12}. 
Many methods have been developed to solve this problem.
Most of the existing methods involve a reconstruction \emph{algorithm} that inputs a pair $(\vec{y},\vec{A})$ and produces a sparse estimate $\hvec{x}$.
A myriad of such algorithms have been proposed,
including both sequential (e.g., greedy) and iterative varieties.
Some relevant algorithms will be reviewed in \secref{algorithms}.

Recently, a different approach to solving this problem has emerged along the lines of ``deep learning'' \cite{Goodfellow:Book:16}, whereby a many-layer \emph{neural network} is optimized to minimize reconstruction mean-squared error (MSE) on a large set of training examples\footnote{%
Since orthonormal $\vec{\Psi}$ implies $\vec{x}=\vec{\Psi}\tran\vec{s}$, training examples of the form $\{(\vec{y}\of{d},\vec{s}\of{d})\}$ can be converted to $\{(\vec{y}\of{d},\vec{x}\of{d})\}_{d=1}^D$ via $\vec{x}\of{d}=\vec{\Psi}\tran\vec{s}\of{d}$.}
$\{(\vec{y}\of{d},\vec{x}\of{d})\}_{d=1}^D$.
Once trained, the network can be used to predict the sparse $\vec{x}\true$ that corresponds to a new input $\vec{y}$.
Although the operator $\vec{A}$ and signal/noise statistics are not explicitly used when training, the learned network will be implicitly dependent on those parameters.
Previous work (e.g.,
\cite{Gregor:ICML:10, 
Sprechmann:ICML:12, 
Kamilov:SPL:16, 
Wang:CVPR:15,Wang:AAAI:16}
)
has shown that the deep-learning approach to solving sparse linear inverse problems has the potential to offer significant improvements, in both accuracy and complexity, over traditional algorithms like ISTA \cite{Chambolle:TIP:98} and FISTA \cite{Beck:JIS:09}.
A short review of relevant concepts from deep learning will be provided in \secref{deep}.

In this paper, we show how recent advances in iterative reconstruction algorithms suggest modifications to traditional neural-network architectures that yield improved accuracy and complexity when solving sparse linear inverse problems.
In particular, we show how ``Onsager correction,'' which lies at the heart of the approximate message passing (AMP) \cite{Donoho:PNAS:09} and vector AMP (VAMP) \cite{Rangan:VAMP} algorithms, can be employed to construct deep networks that i) require fewer layers to reach a given level of accuracy and ii) yield greater accuracy overall.
To our knowledge, the use of Onsager correction in deep networks is novel.

The contributions of our work are as follows.
First, in \secref{lamp}, we show how the soft-thresholding-based AMP algorithm from \cite{Donoho:PNAS:09} can be ``unfolded'' to form a feedforward neural network whose MSE-optimal parameters can be learned using a variant of back-propagation.
The structure of the resulting ``learned AMP'' (LAMP) network is similar to that of learned ISTA (LISTA) \cite{Gregor:ICML:10} but contains additional ``bypass'' paths whose gains are set in a particular way.
While bypass paths can also be found in recently proposed ``residual networks'' \cite{He:ResNet,Veit:NIPS:16} and ``highway networks'' \cite{Srivastava:NIPS:15}, the bypass paths in LAMP have a different topology and a different gain-control mechanism.
We show numerically that LAMP's outputs are more accurate than those of LISTA at each iteration, in some cases by more than a factor of $10$.
To isolate the effects of LAMP's change in network topology, the aforementioned experiments restrict the shrinkage function to classical soft-thresholding.
 
Next, in \secref{lampgen}, we show that the accuracy of LAMP can be significantly improved by learning jointly MSE-optimal shrinkage functions and linear transforms.
In particular, we consider several families of shrinkage functions, each controlled by a small number of learnable parameters:
piecewise linear functions,
exponential shrinkage functions,
cubic B-splines, and
Bernoulli-Gaussian denoisers.
Our work in this section is inspired by \cite{Kamilov:SPL:16}, which learned cubic B-splines for ISTA, but goes farther in that it
i) considers shrinkage families beyond splines, 
ii) jointly learns the shrinkage functions and linear transforms,
and iii) includes Onsager correction.

Then, in \secref{lvamp}, we show how the VAMP algorithm from \cite{Rangan:VAMP} can be unfolded to form a feedforward neural network whose MSE-optimal linear-transforms and shrinkage-functions can be jointly learned using a variant of back-propagation.
Interestingly, we find that learned LVAMP parameters are nearly identical to the prescribed matched-VAMP parameters (i.e., VAMP under statistically matched prior and likelihood) when the signal $\vec{x}$ is i.i.d.
In this sense, matched VAMP ``predicts'' the parameters learned by back-propagation.
Furthermore, since the parameters prescribed by VAMP have an intuitive interpretation based on MMSE estimation principles, VAMP ``explains'' the parameters learned by back-propagation.

Finally, in \secref{5G}, we apply the proposed networks to two problems arising in 5th-generation (5G) wireless communications: 
the \emph{compressive random access} problem 
and the \emph{massive-MIMO channel-estimation} problem.

An early version of this work appeared in \cite{Borgerding:GSIP:16}.
There, we proposed the LAMP-$\ell_1$ algorithm and compared it to LISTA.
In this work, we go beyond \cite{Borgerding:GSIP:16} by 
i) providing justification for our LAMP-$\ell_1$ parameterization (in \appref{redefine}),
ii) jointly optimizing the shrinkage functions and the linear stages of LAMP, 
iii) proposing the LVAMP method, and
iv) detailing two applications to 5G communications.

\subsubsection*{Notation} We use
capital boldface letters like $\vec{A}$ for matrices,
small boldface letters like $\vec{a}$ for vectors,
$(\cdot)\tran$ for transposition,
and $a_n=[\vec{a}]_n$ to denote the $n$th element of $\vec{a}$.
Also, we use
$\|\vec{A}\|_2$ for the spectral norm of $\vec{A}$, 
$\|\vec{a}\|_p=(\sum_n |a_n|^p)^{1/p}$ for the $\ell_p$ norm of $\vec{a}$ when $p>0$,
and
$\|\vec{a}\|_0=|\{a_n:a_n\neq 0\}|$ for the $\ell_0$ or ``counting'' pseudo-norm of $\vec{a}$.
Likewise, we use
$\Diag(\vec{a})$ for the diagonal matrix created from vector $\vec{a}$, 
$\vec{I}_N$ for the $N\times N$ identity matrix
and
$\vec{0}$ for the zero vector. 
For a random vector $\vec{x}$, we denote
its probability density function (pdf) by $p(\vec{x})$
and
its expectation by $\E[\vec{x}]$.
For a random variable $x$, we denote
its variance by $\var[x]$.
Similarly, we use
$p(\cdot|\vec{y})$, $\E[\cdot|\vec{y}]$, and $\var[\cdot|\vec{y}]$ 
for the pdf, expectation, and variance (respectively) conditioned on $\vec{y}$.
We refer to
the Dirac delta pdf using $\delta(\vec{x})$
and to
the pdf of a Gaussian random vector $\vec{x}\in\Real^N$ with mean $\vec{a}$ and covariance $\vec{C}$ using $\mc{N}(\vec{x};\vec{a},\vec{C})=\exp( -(\vec{x}-\vec{a})\tran\vec{C}^{-1}(\vec{x}-\vec{a})/2 )/\sqrt{(2\pi)^N|\vec{C}|}$.
Finally, we use $\sgn(\cdot)$ to denote the signum function, where $\sgn(x)=1$ when $x\geq 0$ and $\sgn(x)=-1$ when $x<0$.

\section{Iterative Algorithms and Deep Learning} \label{sec:background}

\subsection{Iterative Algorithms} \label{sec:algorithms}

One of the best known algorithmic approaches to solving the sparse linear inverse problem is through solving the convex optimization problem \cite{Tibshirani:JRSSb:96,Chen:JSC:98}
\begin{align}
\hvec{x}
&= \arg\min_{\vec{x}} \tfrac{1}{2} \|\vec{y}-\vec{Ax}\|_2^2 + \lambda \|\vec{x}\|_1
\label{eq:lasso} ,
\end{align}
where $\lambda >0$ is a tunable parameter that controls the tradeoff between sparsity and measurement fidelity in $\hvec{x}$.
The convexity of \eqref{lasso} leads to provably convergent algorithms and bounds on the performance of the estimate $\hvec{x}$ (see, e.g., \cite{Candes:CPAM:06}).
In the sequel, we will refer to \eqref{lasso} as the ``$\ell_1$'' problem.

\subsubsection{ISTA} \label{sec:ista}

One of the simplest approaches to solving \eqref{lasso} is the iterative shrinkage/thresholding algorithm (ISTA) \cite{Chambolle:TIP:98}, which iterates the steps (for $t=0,1,2,\dots$ and $\hvec{x}_0=\vec{0}$)
\begin{subequations} \label{eq:ista}
\begin{align}
\vec{v}_t
&= \vec{y} - \vec{A}\hvec{x}_t \\
\hvec{x}_{t+1}
    &= \shrinkage{st} \left( \hvec{x}_t + \beta\vec{A}\tran \vec{v}_t; \lambda \right) ,
\end{align}
\end{subequations}
where $\beta$ is a stepsize,
$\vec{v}_t$ is the iteration-$t$ residual measurement error, and
$\shrinkage{st}\left(\cdot;\lambda\right):\Real^N\rightarrow\Real^N$ is the ``soft thresholding'' shrinkage function, defined componentwise as
\begin{align}
    [\shrinkage{st}\left(\vec{r};\lambda\right)]_j
&\defn \sgn(r_j) \max\{|r_j|-\lambda,0\}
\label{eq:soft_thresh} .
\end{align}

\subsubsection{FISTA} \label{sec:fista}

Although ISTA is guaranteed to converge under $\beta\in(0,1/\|\vec{A}\|^2_2)$ \cite{Daubechies:CPAM:04}, it converges somewhat slowly and so many modifications have been proposed to speed it up.
Among the most famous is ``fast ISTA'' (FISTA) \cite{Beck:JIS:09},
\begin{subequations} \label{eq:fista}
\begin{align}
\vec{v}_t
&= \vec{y} - \vec{A}\hvec{x}_t \\
\hvec{x}_{t+1}
    &= \shrinkage{st} \left(
   \hvec{x}_t
   + \beta\vec{A}\tran  \vec{v}_{t}
   + \tfrac {t-2} {t+1} \left( \hvec{x}_t - \hvec{x}_{t-1} \right)
   ; \lambda \right) ,
\end{align}
\end{subequations}
which converges in roughly an order-of-magnitude fewer iterations than ISTA (see \figref{ista_fista_amp_vamp_Giid_k15}).

\subsubsection{AMP} \label{sec:amp}

Recently, an approximate message passing (AMP) algorithm \cite{Donoho:PNAS:09,Montanari:Chap:12} was proposed for the $\ell_1$ problem. 
The resulting algorithm, which we call AMP-$\ell_1$, manifests as
\begin{subequations} \label{eq:amp}
\begin{align}
\vec{v}_t
&= \vec{y} - \vec{A}\hvec{x}_t + b_t \vec{v}_{t-1}
\label{eq:amp1} \\
\hvec{x}_{t+1}
&= \shrinkage{st} \left( \hvec{x}_t + \vec{A}\tran\vec{v}_t; \lambda_t \right),
\label{eq:amp2}
\end{align}
\end{subequations}
where $\hvec{x}_0=\vec{0}$, $\vec{v}_{-1}=\vec{0}$, $t\in\{0,1,2,\dots\}$, and
\begin{align}
b_t
&= \tfrac{1}{M}\|\hvec{x}_t\|_0
\label{eq:bt} \\
\lambda_t
&= \tfrac{\alpha}{\sqrt{M}} \norm{\vec{v}_t}_2
\label{eq:lambda} .
\end{align}
In \eqref{lambda}, $\alpha$ is a tuning parameter that has a one-to-one correspondence with $\lambda$ in \eqref{lasso} \cite{Montanari:Chap:12}.
Comparing AMP-$\ell_1$ to ISTA, we see two major differences:
i) AMP's residual $\vec{v}_t$ in \eqref{amp1} includes the ``Onsager correction'' term $b_t\vec{v}_{t-1}$, and
ii) AMP's shrinkage threshold $\lambda_t$ in \eqref{amp2} takes the prescribed, $t$-dependent value \eqref{lambda}.
In the sequel, we explain the rationale behind these differences.

AMP can in fact be used with any Lipschitz-continuous shrinkage function.
For this, we write the AMP algorithm as 
\begin{subequations} \label{eq:ampgen}
\begin{align}
\vec{v}_t
&= \vec{y} - \vec{A}\hvec{x}_t + b_t \vec{v}_{t-1} 
\label{eq:ampgen1} \\
\hvec{x}_{t+1}
&= \shrinkage{} \left( \hvec{x}_t + \vec{A}\tran\vec{v}_t; \sigma_t, \vec{\theta}_t \right) 
\label{eq:ampgen2} ,
\end{align}
\end{subequations}
where $\hvec{x}_0=\vec{0}$, $\vec{v}_{-1}=\vec{0}$, $t\in\{0,1,2,\dots\}$, and
\begin{align}
b_{t+1}
&= \frac{1}{M} \sum_{j=1}^N \left. \frac{\partial [\vec{\eta}(\vec{r};\sigma_t,\vec{\theta}_t)]_j}{\partial r_j} \right|_{\vec{r}=\hvec{x}_t + \vec{A}\tran\vec{v}_t}
\label{eq:btgen} \\
\sigma_t^2
&= \tfrac{1}{M}\|\vec{v}_t\|_2^2 .
\label{eq:sigma}
\end{align}
In writing \eqref{ampgen2}, we assume that the shrinkage function $\vec{\eta}$ accepts the noise-standard-deviation estimate $\sigma_t$ as an argument. 
Although this is not a required feature of AMP, we find it useful in the sequel.
It is straightforward to show that AMP in \eqref{ampgen}-\eqref{sigma} reduces to AMP-$\ell_1$ from \eqref{amp}-\eqref{lambda} when $\vec{\eta}(\vec{r}_t;\sigma_t,\alpha)=\shrinkage{st}(\vec{r}_t;\alpha\sigma_t)$ and $\vec{\theta}_t=\alpha$. 

When $\vec{A}$ is a typical realization of a large i.i.d.\ sub-Gaussian random matrix with variance-$M^{-1}$ entries and $\vec{\eta}(\cdot)$ has identical scalar components, the Onsager correction \emph{decouples} the AMP iterations in the sense that the input to the shrinkage function,
\begin{align}
\vec{r}_t
\defn \hvec{x}_t + \vec{A}\tran\vec{v}_t ,
\end{align}
can be accurately modeled as\footnote{The AMP model \eqref{lsl}-\eqref{sigma} is provably accurate in the large-system limit (i.e., $M,N\rightarrow\infty$ with $M/N$ converging to a positive constant) \cite{Bayati:TIT:11,Bayati:AAP:15}.}
\begin{align}
\vec{r}_t
&= \vec{x}\true + \mc{N}(\vec{0},\sigma_t^2\vec{I}_N)
\label{eq:lsl}
\end{align}
with $\sigma_t^2$ from \eqref{sigma}.
In other words, the Onsager correction ensures that the shrinkage input is an AWGN-corrupted version of the true signal $\vec{x}\true$ with known variance $\sigma_t^2$.
(See \figref{qqplots}(b) for numerical evidence.)
The resulting ``denoising'' problem, that of estimating $\vec{x}\true$ from $\vec{r}_t$, is well understood.

For example, when the elements of $\vec{x}\true$ are statistically independent with known prior $p(\vec{x})=\prod_{j=1}^N p_j(x_j)$, the MSE-optimal denoiser\footnote{AMP with MSE-optimal denoising was first described in \cite{Donoho:ITW:10a}.} is simply the posterior mean estimator
(i.e., $\hat{x}_{t+1,j}=\E\{x_j|r_{t,j};\sigma_t\}$),
which can be computed in closed form for many distributions $p_j(\cdot)$.
In the case that $p_j(\cdot)$ are unknown, we may be more interested in the \emph{minimax} denoiser, i.e., the minimizer of the maximum MSE over an assumed family of priors.
Remarkably, for generic sparse priors, i.e., $p_j(x_j) = (1-\gamma)\delta(x_j)+\gamma \tilde{p}_j(x_j)$ with $\gamma\in(0,1)$ and arbitrary unknown $\tilde{p}_j(\cdot)$,
soft-thresholding \eqref{soft_thresh} with a threshold proportional to the AWGN standard deviation (i.e., $\lambda_t=\alpha\sigma_t$ recalling \eqref{sigma}) is nearly minimax optimal \cite{Montanari:Chap:12}.
Thus, we can interpret the AMP-$\ell_1$ algorithm \eqref{amp} as a nearly minimax approach to the sparse linear inverse problem under unknown $\tilde{p}_j(\cdot)$

The behavior of AMP is well understood when $\vec{A}$ is i.i.d.\ sub-Gaussian \cite{Bayati:TIT:11,Bayati:AAP:15}, but even small deviations from this model can lead AMP to diverge \cite{Vila:ICASSP:15} or at least behave in ways that are not well understood.

\subsubsection{Vector AMP} \label{sec:vamp}

Very recently, the VAMP algorithm (see \algref{vamp})
was proposed in \cite{Rangan:VAMP} to address AMP's fragility with respect to the matrix $\vec{A}$.
The VAMP algorithm retains all the desirable properties of the original AMP (i.e.,
low per-iteration complexity, 
very few iterations to convergence,
and shrinkage inputs $\vec{r}_t$ that obey the AWGN model \eqref{lsl}),
but over a much larger class of matrices: those that are
large and right-rotationally invariant $\vec{A}$.

A \emph{right-rotationally invariant} matrix $\vec{A}$ is a random matrix whose distribution remains the same after right multiplication by any fixed orthogonal matrix. 
An intuitive understanding of such matrices arises from their singular value decomposition (SVD).
Suppose that 
\begin{align}
\vec{A}
&= \vec{USV}\tran
\label{eq:svd}
\end{align}
is the economy-sized\footnote{By ``economy-sized,'' we mean that if
$R\defn\rank(\vec{A})$
and $\vec{s}\in\Real_+^R$ contains the positive singular values of $\vec{A}$,
then 
$\vec{S}=\Diag(\vec{s})\in\Real^{R\times R}$,
$\vec{U}\tran\vec{U}=\vec{I}_R$, and
$\vec{V}\tran\vec{V}=\vec{I}_R$.}
SVD of $\vec{A}\in\Real^{M\times N}$.
For right-rotationally invariant $\vec{A}$, the matrix
$\vec{V}$ will contain the first $R$ columns of a matrix that is uniformly distributed on the group of $N\times N$ orthogonal matrices.
Note that i.i.d.\ Gaussian matrices are a special case of right-rotationally invariant, one where $\vec{U}$ is random orthogonal and $\vec{s}$ has a particular distribution. 
Importantly, VAMP behaves well under \emph{any} orthogonal matrix $\vec{U}$ and \emph{any} singular values $\vec{s}$, as long as the dimensions $M,N$ are large enough \cite{Rangan:VAMP}.

\begin{algorithm}[t]
\caption{Vector AMP \cite{Rangan:VAMP}}
\begin{algorithmic}[1]  \label{alg:vamp}
\REQUIRE{
LMMSE estimator $\tvec{\eta}(\cdot;\tilde{\sigma},\tvec{\theta})$ from \eqref{lmmse},
shrinkage $\vec{\eta}(\cdot;\sigma,\vec{\theta})$,
max iterations $T$, 
parameters $\{\vec{\theta}_t\}_{t=1}^T$ and $\tvec{\theta}$.}
\STATE{Select initial $\tvec{r}_1$ and $\tilde{\sigma}_1>0$.}
\FOR{$t=1,2,\dots,T$}
    \STATE{// LMMSE stage:}
    \STATE{$\tvec{x}_t = \tvec{\eta}\big(\tvec{r}_t;\tilde{\sigma}_t,\tvec{\theta}\big)$}
        \label{line:xtil}
    \STATE{$\tilde{\dvg}_t = \big\langle \tvec{\eta}'\big(\tvec{r}_t;\tilde{\sigma}_t,\tvec{\theta}\big) \big\rangle$}
        \label{line:atil}
    \STATE{$\vec{r}_t = (\tvec{x}_t - \tilde{\dvg}_t\tvec{r}_t)/(1-\tilde{\dvg}_t)$}
        \label{line:r}
    \STATE{$\sigma_t^2 = \tilde{\sigma}_t^2\tilde{\dvg}_t/(1-\tilde{\dvg}_t)$}
        \label{line:sig}
\vspace{1mm}
    \STATE{// Shrinkage stage:}
    \STATE{$\hvec{x}_t = \vec{\eta}(\vec{r}_t;\sigma_t,\vec{\theta}_t)$}
        \label{line:x}
    \STATE{$\dvg_t = \bkt{ \vec{\eta}'(\vec{r}_t,\sigma_t,\vec{\theta}_t) }$}
        \label{line:a}
    \STATE{$\tvec{r}_{t+1} = (\hvec{x}_t - \dvg_t\vec{r}_t)/(1-\dvg_t)$}
        \label{line:rtil}
    \STATE{$\tilde{\sigma}_{t+1}^2 = \sigma_t^2\dvg_t/(1-\dvg_t)$}
        \label{line:sigtil}
\ENDFOR
\STATE{Return $\hvec{x}_{T}$.}
\end{algorithmic}
\end{algorithm}

The VAMP algorithm is defined in \algref{vamp}.
The algorithm can be seen to consist of two stages, 
each comprising the same four steps:
estimation (lines~\ref{line:xtil} and \ref{line:x}), 
divergence computation (lines~\ref{line:atil} and \ref{line:a}),
Onsager correction (lines~\ref{line:r} and \ref{line:rtil}), 
and variance computation (lines~\ref{line:sig} and \ref{line:sigtil}).
The only difference between the two stages is their choice of estimator.
The first stage uses
\begin{align}
\lefteqn{
\tvec{\eta}\big(\tvec{r}_t;\tilde{\sigma}_t,\tvec{\theta}\big)
} \label{eq:lmmse}\\
&\defn \vec{V}\left(\Diag(\vec{s})^2+\frac{\sigma_w^2}{\tilde{\sigma}_t^2}\vec{I}_R\right)^{-1}\left(\Diag(\vec{s})\vec{U}\tran\vec{y}+\frac{\sigma_w^2}{\tilde{\sigma}_t^2}\vec{V}\tran\tvec{r}_t\right)
\nonumber ,
\end{align}
which depends on the measurements $\vec{y}$ and the parameters
\begin{align}
\tvec{\theta} \defn \{\vec{U},\vec{s},\vec{V},\sigma_w\} 
\label{eq:thetatil} ,
\end{align}
while the second stage performs componentwise nonlinear shrinkage via
$\vec{\eta}(\vec{r}_t;\sigma_t,\vec{\theta}_t)$,
just as in step \eqref{ampgen2} of the AMP algorithm.

Lines~\ref{line:atil} and \ref{line:a} in \algref{vamp} compute the average of the diagonal entries of the Jacobian
of $\tvec{\eta}(\cdot;\tilde{\sigma}_t,\tvec{\theta})$ and $\vec{\eta}(\cdot;\sigma_t,\vec{\theta}_t)$, respectively. 
That is,
\begin{align}
\bkt{\vec{\eta}'(\vec{r};\sigma,\vec{\theta})}
&\defn \frac{1}{N}\sum_{j=1}^N \frac{\partial [\vec{\eta}(\vec{r};\sigma,\vec{\theta})]_j}{\partial r_j}
\label{eq:divergence} .
\end{align}
From \eqref{lmmse}, we see that the Jacobian of $\tvec{\eta}(\cdot;\tilde{\sigma}_t,\tvec{\theta})$ is
\begin{align}
\frac{\sigma_w^2}{\tilde{\sigma}_t^2}
\vec{V}\left(\Diag(\vec{s})^2+\frac{\sigma_w^2}{\tilde{\sigma}_t^2}\vec{I}_R\right)^{-1} \vec{V}\tran ,
\end{align}
and so the average of its diagonal (or $N^{-1}$ times its trace) is
\begin{align}
\big\langle
\tvec{\eta}'\big(\tvec{r}_t;\tilde{\sigma}_t,\tvec{\theta}\big)
\big\rangle
&= \frac{1}{N}\sum_{i=1}^R \frac{1}{s_i^2 \tilde{\sigma}_t^2 / \sigma_w^2 + 1} 
\label{eq:div_lmmse} .
\end{align}

The first-stage estimator $\tvec{\eta}(\cdot;\tilde{\sigma}_t,\tvec{\theta})$ in \eqref{lmmse} can be interpreted as computing the MMSE estimate of $\vec{x}\true$ under the likelihood function
\begin{align}
p(\vec{y}|\vec{x}\true)
&= \mc{N}(\vec{y};\vec{Ax}\true,\sigma_w^2\vec{I})
\label{eq:like} ,
\end{align}
which follows from \eqref{y} under the assumption that $\vec{w}\sim\mc{N}(\vec{0},\sigma_w^2\vec{I})$
and the pseudo-prior
\begin{align}
\vec{x}\true &\sim \mc{N}(\tvec{r}_t,\tilde{\sigma}_t^2\vec{I})
\label{eq:pseudoprior} .
\end{align}
We refer to \eqref{pseudoprior} as a ``pseudo'' prior because it is constructed internally by VAMP at each iteration $t$.
The MMSE estimate of $\vec{x}$ is then given by the conditional mean $\E\{\vec{x}|\vec{y}\}$, which in the case of \eqref{like}-\eqref{pseudoprior} is
\begin{align}
\left(\vec{A}\tran\vec{A} + \frac{\sigma_w^2}{\tilde{\sigma}_t^{2}}\vec{I}_N\right)^{-1}
        \left(\vec{A}\tran\vec{y} + \frac{\sigma_w^2}{\tilde{\sigma}_t^{2}}\tvec{r}_t \right)
\label{eq:lmmse2} .
\end{align}
Replacing $\vec{A}$ in \eqref{lmmse2} with its SVD from \eqref{svd} yields the expression in \eqref{lmmse}.
Since the estimate is linear in $\tvec{r}_t$, we refer to the first stage as the ``linear MMSE'' (LMMSE) stage.

The 2nd-stage estimator $\vec{\eta}(\cdot;\sigma_t,\vec{\theta}_t)$, in \lineref{x} of \algref{vamp}, essentially denoises the pseudo-measurement 
\begin{align}
\vec{r}_t
&= \vec{x}\true + \mc{N}(\vec{0},\sigma_t^2) 
\label{eq:lsl2} .
\end{align}
The AWGN-corruption model \eqref{lsl2} holds under large, right-rotationally invariant $\vec{A}$ and $\vec{\eta}(\cdot)$ with identical components, as proven in \cite{Rangan:VAMP}.
If the prior $p(\vec{x}\true)$ on $\vec{x}\true$ was known,\footnote{Although the prior and noise variance are often unknown in practice, they can be learned online using the EM-VAMP approach from \cite{Fletcher:ICASSP:17}.} then it would be appropriate to choose the MMSE denoiser for $\vec{\eta}$:
\begin{align}
\vec{\eta}(\vec{r}_t;\sigma_t,\vec{\theta}_t)
&= \E\{\vec{x}\true | \vec{r}_t\} ,
\end{align}
With an i.i.d.\ signal and MMSE denoiser, VAMP produces a sequence $\{\hvec{x}_t\}$ whose fixed points have MSE consistent with the replica prediction of MMSE from \cite{Tulino:TIT:13}.
In the sequel, we shall refer to VAMP with MMSE i.i.d.-signal denoising and known $\sigma_w^2$ as ``matched VAMP.''

In summary, VAMP alternates between 
i) MMSE inference of $\vec{x}\true$ under 
likelihood $\mc{N}(\vec{y};\vec{Ax}\true,\sigma_w^2\vec{I})$ and 
pseudo-prior $\mc{N}(\vec{x}\true;\tvec{r}_t,\tilde{\sigma}_t^2\vec{I})$, and
ii) MMSE inference of $\vec{x}\true$ under 
pseudo-likelihood $\mc{N}(\vec{r}_t;\vec{x}\true,\sigma_t^2\vec{I})$ and
prior $\vec{x}\true\sim p(\vec{x}\true)$.
The intermediate quantities 
$\tvec{r}_t$ and $\vec{r}_t$
are updated in each stage of VAMP using the Onsager correction terms
$-\dvg_t\vec{r}_t$ and $-\tilde{\dvg}_t\tvec{r}_t$, respectively,
where $\dvg_t$ and $\tilde{\dvg}_t$ are the divergences\footnote{%
Notice that the Onsager correction term $b_{t+1}\vec{v}_t$ in AMP step \eqref{ampgen1} also involves a ($N/M$-scaled) divergence, $b_{t+1}$, defined in \eqref{btgen}.
} 
associated with the estimators $\vec{\eta}$ and $\tvec{\eta}$.
Essentially, the Onsager correction acts to decouple the two stages (and iterations) of VAMP from each other so that local MSE optimization at each stage leads to global MSE optimization of the algorithm.

\subsubsection{Comparison of ISTA, FISTA, AMP-\texorpdfstring{$\ell_1$}{L1}, and VAMP-\texorpdfstring{$\ell_1$}{L1}} \label{sec:compare}

For illustration, we now compare the average per-iteration behavior of ISTA, FISTA, AMP-$\ell_1$, and VAMP-$\ell_1$
in two scenarios: 
i) for an $\vec{A}$ drawn i.i.d.\ $\mc{N}(0,M^{-1})$, and
ii) when the singular values of the same $\vec{A}$ are replaced by a geometric series that yields the condition number $\kappa(\vec{A})=15$.
That is, $s_i/s_{i-1}=\rho~\forall i>1$, with $\rho$ set to achieve the condition-number $s_1/s_M=15$ and $s_1$ set to yield $\|\vec{A}\|_F^2=N$.
In both cases,
the problem dimensions were $N=500$ and $M=250$;
the elements of $\vec{x}\true$ were i.i.d.\ $\mc{N}(0,1)$ with probability $\gamma=0.1$ and were otherwise set to zero (i.e., $\vec{x}\true$ was Bernoulli-Gaussian); and
the noise $\vec{w}$ was i.i.d.\ $\mc{N}(0,\sigma_w^2)$, with $\sigma_w^2$ set to yield a signal-to-noise ratio (SNR) $\E\{\|\vec{Ax}\true\|^2\}/\E\{\|\vec{w}\|^2\}$ of $40$~dB.
Recall that ISTA, FISTA, AMP-$\ell_1$, and VAMP-$\ell_1$ all estimate $\vec{x}$ by iteratively minimizing \eqref{lasso} for a chosen value of $\lambda$ (selected via $\alpha$ in the case of AMP and VAMP).
We chose the minimax optimal value of $\alpha$ for AMP (which equals $1.1402$ since $\gamma=0.1$ \cite{Montanari:Chap:12}) and VAMP, and we used the corresponding $\lambda$ for ISTA and FISTA.

\Figref{ista_fista_amp_vamp_Giid_k15} shows the average normalized MSE (NMSE) versus iteration $t$, where NMSE$_t \defn \|\hvec{x}_t-\vec{x}\true\|_2^2/\|\vec{x}\true\|_2^2$ and $1000$ realizations of $(\vec{x},\vec{w})$ were averaged.
In \figref{ista_fista_amp_vamp_Giid_k15}(a), we see that AMP-$\ell_1$ required an order-of-magnitude fewer iterations than FISTA, which required an order-of-magnitude fewer iterations than ISTA.  
Meanwhile, we see that VAMP-$\ell_1$ required about half the iterations of AMP-$\ell_1$.
In \figref{ista_fista_amp_vamp_Giid_k15}(b), AMP-$\ell_1$ is not shown because it diverged.  
But VAMP-$\ell_1$ required an order-of-magnitude fewer iterations than FISTA, which required an order-of-magnitude fewer iterations than ISTA.

\putFrag{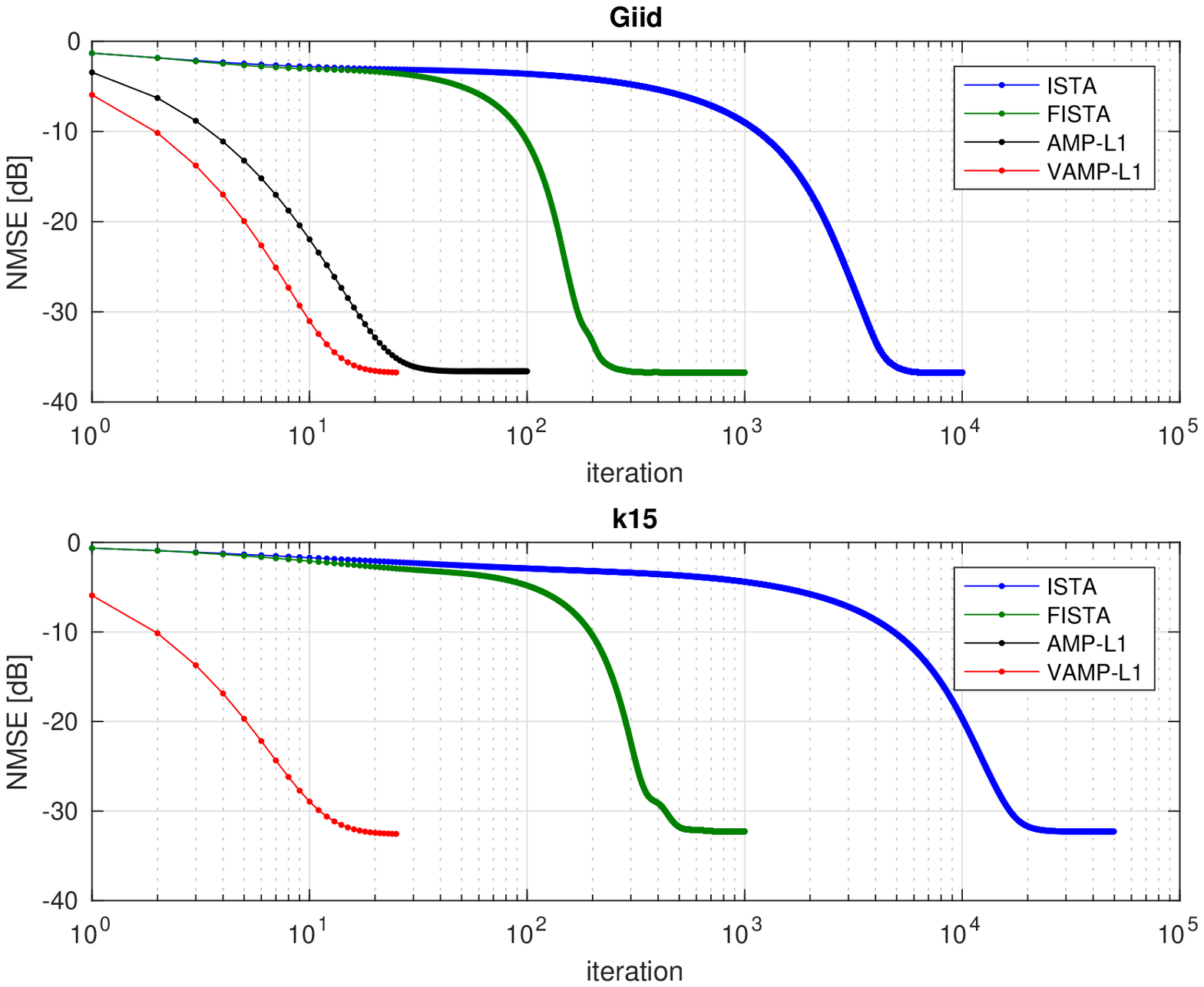}
   {Average NMSE versus iteration for VAMP-$\ell_1$, AMP-$\ell_1$, FISTA, and ISTA under (a) i.i.d.\ Gaussian $\vec{A}$ and (b) $\vec{A}$ with condition number $\kappa=15$.  Note that the horizontal axis is plotted on a log scale.}
   {\figsize}
   {\psfrag{ISTA}[l][l][0.5]{\sf ISTA}
    \psfrag{FISTA}[l][l][0.5]{\sf FISTA}
    \psfrag{AMP-L1}[l][l][0.5]{\sf AMP-$\ell_1$}
    \psfrag{VAMP-L1}[l][l][0.5]{\sf VAMP-$\ell_1$}
    \psfrag{Recovery NMSE (dB)}[B][][0.7]{\sf average NMSE [dB]}
    \psfrag{iteration}[B][][0.7]{\sf iteration}
    \psfrag{NMSE [dB]}[][][0.7]{\sf average NMSE [dB]}
    \psfrag{iidG}[][][0.7]{\sf (a)}
    \psfrag{kappa=15}[][][0.7]{\sf (b)}
    \psfrag{Giid}{}
    \psfrag{k15}{}
   }

\subsection{Deep Learning} \label{sec:deep}

In deep learning \cite{Goodfellow:Book:16}, training data $\{(\vec{y}\of{d},\vec{x}\of{d})\}_{d=1}^D$ comprised of (feature,label) pairs are used to train the parameters of a deep neural network, with the goal of accurately predicting the unknown label $\vec{x}\true$ associated with a newly observed feature $\vec{y}$.
The deep network accepts $\vec{y}$ and subjects it to many layers of processing, where each layer usually consists of a linear transformation followed by a simple, componentwise nonlinearity.

Typically, the label space is discrete (e.g., $\vec{y}$ is an image and $\vec{x}$ is its class in \{cat, dog, \dots, tree\}).
In our sparse linear inverse problem, however, the ``labels'' $\vec{x}$ are continuous and high-dimensional.
Remarkably, Gregor and LeCun demonstrated in \cite{Gregor:ICML:10} that a well-constructed deep network can accurately predict even labels such as these.

The neural network architecture proposed in \cite{Gregor:ICML:10} is closely related to the ISTA algorithm discussed in \secref{ista}.
To understand the relation, we rewrite the ISTA iteration \eqref{ista} as
\begin{align}
\hvec{x}_{t+1}
    &= \shrinkage{st} \left( \vec{S}\hvec{x}_t + \vec{B}\vec{y}; \lambda \right)
\text{~~with~}
\begin{cases}
\vec{B}= \beta\vec{A}\tran \\
\vec{S}= \vec{I}_N-\vec{B}\vec{A}
\end{cases}
\label{eq:ista2}
\end{align}
and ``unfold'' the iterations $t=1,\dots,T$, resulting in the $T$-layer feed-forward neural network shown in \figref{lista_3layer}.

\putFrag{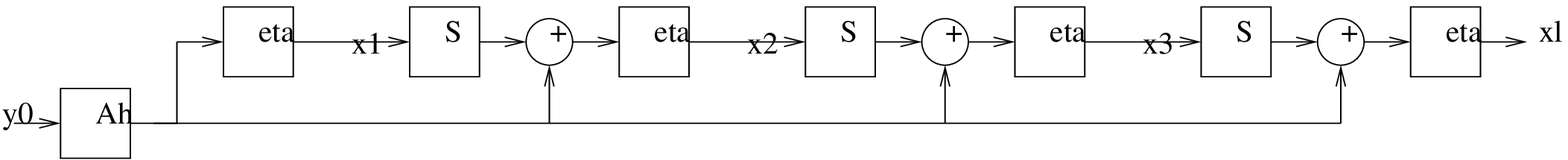}
        {The feed-forward neural network constructed by unfolding $T\!=\!4$ iterations of ISTA.}
        {3.4}
        {\newcommand{\sz}{0.75}
         \newcommand{\szz}{0.75}
         \psfrag{+}[c][Bl][\szz]{$+$}
         \psfrag{-}[r][Bl][\szz]{$-$}
         \psfrag{x}[c][Bl][\sz]{$\times$}
         \psfrag{y0}[r][Bl][\sz]{$\vec{y}$}
         \psfrag{0}[r][Bl][\sz]{$\vec{0}$}
         \psfrag{r}[t][Bl][\sz]{$\vec{r}_t$}
         \psfrag{Ah}[c][Bl][\sz]{$\vec{B}$}
         \psfrag{S}[c][Bl][\sz]{$\vec{S}$}
         \psfrag{x1}[t][Bl][\szz]{$\hvec{x}_1$}
         \psfrag{x2}[t][Bl][\szz]{$\hvec{x}_2$}
         \psfrag{x3}[t][Bl][\szz]{$\hvec{x}_3$}
         \psfrag{xl}[l][Bl][\sz]{$\hvec{x}_4$}
         \psfrag{eta}[c][Bl][\sz]{$\shrinkage{st}$}
        }

Whereas ISTA uses the values of $\vec{S}$ and $\vec{B}$ prescribed in \eqref{ista2} and a common value of $\lambda$ at all layers, Gregor and LeCun \cite{Gregor:ICML:10} proposed to use layer-dependent thresholds $\vec{\lambda}\defn[\lambda_1,\lambda_2,\dots,\lambda_T]$ and ``learn'' both the thresholds $\vec{\lambda}$ and the matrices $\vec{B},\vec{S}$ from the training data $\{(\vec{y}\of{d},\vec{x}\of{d})\}_{d=1}^D$ by minimizing the quadratic loss
\begin{align}
\mc{L}_T(\vec{\Theta}) = \frac{1}{D}\sum_{d=1}^D \big\|\hvec{x}_T(\vec{y}\of{d};\vec{\Theta})-\vec{x}\of{d}\big\|_2^2 .
\label{eq:loss}
\end{align}
Here, $\vec{\Theta}=[\vec{B},\vec{S},\vec{\lambda}]$ denotes the set of learnable parameters and $\hvec{x}_T(\vec{y}\of{d};\vec{\Theta})$ the output of the $T$-layer network with input $\vec{y}\of{d}$ and parameters $\vec{\Theta}$.
The resulting approach was coined ``learned ISTA'' (LISTA).

The LISTA \emph{network} generated estimates of comparable MSE with significantly fewer matrix-vector multiplications than existing \emph{algorithms} for the $\ell_1$ problem \eqref{lasso} with optimally tuned regularization parameters (e.g., $\lambda$ or $\alpha$). 
As an example, for the
i.i.d.\ Gaussian version of the
problem described in \secref{compare},
LISTA took only $16$ layers to reach an NMSE of $-35$~dB, whereas 
AMP-$\ell_1$ took $25$ iterations,\footnote{The computational complexity of one layer of LISTA is essentially equal to one iteration of ISTA, FISTA, or AMP.} 
FISTA took $216$, 
and ISTA took $4402$. 
(More details will be given in \secref{numerical1}.)

Other authors have also applied ideas from deep learning to the sparse linear inverse problem.
For example, \cite{Sprechmann:ICML:12} extended the LISTA approach \cite{Gregor:ICML:10} to handle structured sparsity and dictionary learning (when the training data are $\{\vec{y}\of{d}\}_{d=1}^D$ and $\vec{A}$ is unknown).
More recently, \cite{Wang:CVPR:15,Wang:AAAI:16} extended LISTA from the $\ell_2\!+\!\ell_1$ objective of \eqref{lasso} to the $\ell_2\!+\!\ell_0$ objective,
and \cite{Kamilov:SPL:16} proposed to learn the MSE-optimal scalar shrinkage functions $\shrinkage{}$ by learning the parameters of a B-spline.
It has also been proposed to recover signals using deep networks other than the ``unfolded'' type. 
For example, convolutional neural networks and stacked denoising autoencoders
have been applied to
speech enhancement \cite{Hershey:Tech:14},
image denoising \cite{Burger:CVPR:12},
image deblurring \cite{Schuler:CVPR:13,Schmidt:CVPR:14},
image super resolution \cite{Dong:TPAMI:16},
3D imaging \cite{Xin:NIPS:16},
compressive imaging \cite{Mousavi:ALL:15,Kulkarni:CVPR:16,Mousavi:ICASSP:17},
and
video compressive sensing \cite{Iliadis:16}.

\section{Learned AMP-\texorpdfstring{$\ell_1$}{L1}} \label{sec:lamp}

As described earlier, LISTA learns the value of the linear transform $\vec{S}\in\Real^{N\times N}$ that minimizes MSE on the training data.
As noted in \cite{Gregor:ICML:10}, however, the LISTA's performance does not degrade after imposing the structure
\begin{align}
\vec{S}
&=\vec{I}_N-\vec{BA}, 
\label{eq:S}
\end{align}
where $\vec{B}\in\Real^{N\times M}$ and $\vec{A}\in\Real^{M\times N}$, as suggested by \eqref{ista2}.
Since the form of $\vec{S}$ in \eqref{S} involves $2MN$ free parameters, it is advantageous (in memory and training) over unstructured $\vec{S}$ when $M\!<\!N/2$, which is often the case in compressive sensing. 
The structured $\vec{S}$ from \eqref{S} leads to 
network layers of the form shown in \figref{lista_1layer}, with first-layer inputs $\hvec{x}_0=\vec{0}$ and $\vec{v}_0=\vec{y}$.

Although not considered in \cite{Gregor:ICML:10}, the network in \figref{lista_1layer} allows both $\vec{A}$ and $\vec{B}$ to vary with the layer $t$,
allowing for a modest performance improvement (as will be demonstrated in \secref{numerical1}) at the expense of a $T$-fold increase in memory and training complexity.
We will refer to networks that use 
fixed $\vec{A}$ and $\vec{B}$ over all layers $t$ as ``tied,'' and
those that allow $t$-dependent $\vec{A}_t$ and $\vec{B}_t$ as ``untied.''

\putFrag{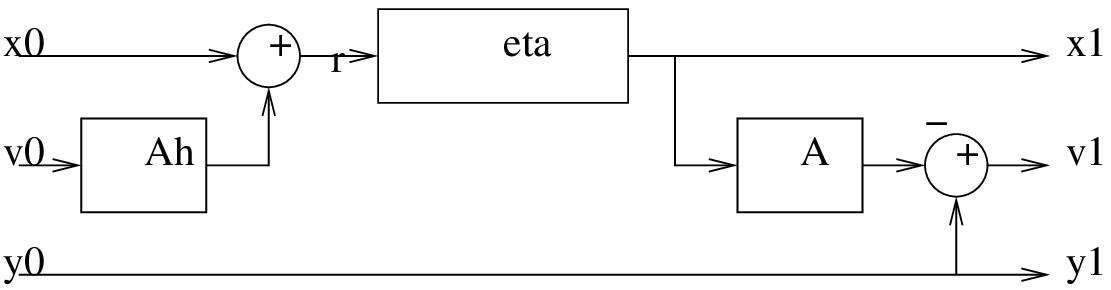}
{The $t$th layer of the LISTA network, with learnable parameters $\vec{A}_t,\vec{B}_t$, and $\lambda_t$.}
        {2.3}
        {\newcommand{\sz}{1.0}
         \newcommand{\szz}{0.8}
         \newcommand{\szzz}{0.7}
         \psfrag{+}[c][Bl][\szz]{$+$}
         \psfrag{-}[r][Bl][\szz]{$-$}
         \psfrag{x}[c][Bl][\sz]{$\times$}
         \psfrag{x0}[r][Bl][\sz]{$\hvec{x}_t$}
         \psfrag{x1}[l][Bl][\sz]{$\hvec{x}_{t\!+\!1}$}
         \psfrag{v0}[r][Bl][\sz]{$\vec{v}_t$}
         \psfrag{v1}[l][Bl][\sz]{$\vec{v}_{t\!+\!1}$}
         \psfrag{y0}[r][Bl][\sz]{$\vec{y}$}
         \psfrag{y1}[l][Bl][\sz]{$\vec{y}$}
         \psfrag{r}[t][Bl][\szz]{$\vec{r}_t$}
         \psfrag{Ah}[c][Bl][\sz]{$\vec{B}_t$}
         \psfrag{A}[c][Bl][\sz]{$\vec{A}_t$}
         \psfrag{eta}[c][Bl][\szz]{$\shrinkage{st}\left(\vec{r}_t;\lambda_t\right)$}
        }

\subsection{The LAMP-\texorpdfstring{$\ell_1$}{L1} Network} \label{sec:lampnet}

We propose to construct a neural network by unfolding the iterations of AMP-$\ell_1$ from \eqref{amp}.
We then propose to learn the MSE-optimal values of the network parameters,
$\{\vec{A}_t,\vec{B}_t,\alpha_t\}_{t=0}^{T-1}$, from training data $\{(\vec{y}\of{d},\vec{x}\of{d})\}_{d=1}^D$.
We will refer to this approach as ``learned AMP-$\ell_1$'' (LAMP-$\ell_1$).
The hope is that it will require fewer layers than LISTA to yield an accurate reconstruction, just as AMP-$\ell_1$ requires many fewer iterations than ISTA to do the same (when $\vec{A}$ is drawn i.i.d.\ Gaussian).

\Figref{lamp_1layer} shows one layer of the LAMP-$\ell_1$ network.
Comparing LAMP-$\ell_1$ to LISTA, we see two main differences:
\begin{enumerate}
\item
LAMP-$\ell_1$ includes a ``bypass'' path from $\vec{v}_t$ to $\vec{v}_{t+1}$ that is not present in LISTA.
This path implements an ``Onsager correction'' whose goal is to \emph{decouple} the layers of the network, just as it decoupled the iterations of the AMP algorithm (recall \secref{amp}).
\item
LAMP-$\ell_1$'s $t$th shrinkage threshold $\lambda_t=\alpha_t\|\vec{v}_t\|_2/\sqrt{M}$ varies with the realization $\vec{v}_t$, whereas LISTA's does not.
\end{enumerate}

\putFrag{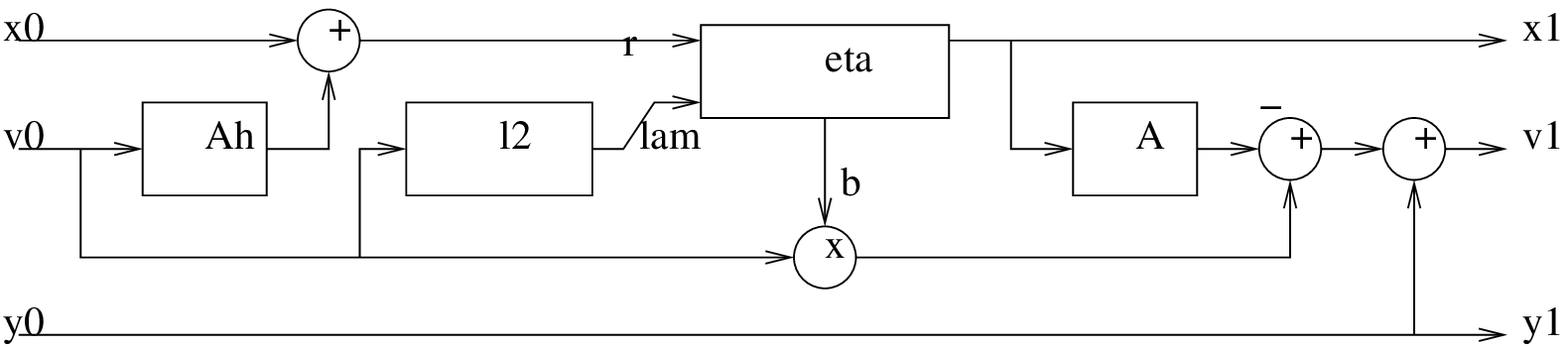}
        {The $t$th layer of the LAMP-$\ell_1$ network, with learnable parameters $\vec{A}_t,\vec{B}_t$, and $\alpha_t$.}
        {3.15}
        {\newcommand{\sz}{1.0}
         \newcommand{\szz}{0.9}
         \newcommand{\szzz}{0.8}
         \psfrag{+}[c][Bl][\sz]{$+$}
         \psfrag{-}[r][Bl][\szz]{$-$}
         \psfrag{x}[c][Bl][\sz]{$\times$}
         \psfrag{x0}[r][Bl][\sz]{$\hvec{x}_t$}
         \psfrag{x1}[l][Bl][\sz]{$\hvec{x}_{t\!+\!1}$}
         \psfrag{v0}[r][Bl][\sz]{$\vec{v}_t$}
         \psfrag{v1}[l][Bl][\sz]{$\vec{v}_{t\!+\!1}$}
         \psfrag{y0}[r][Bl][\sz]{$\vec{y}$}
         \psfrag{y1}[l][Bl][\sz]{$\vec{y}$}
         \psfrag{r}[t][Bl][\szzz]{$\vec{r}_t$}
         \psfrag{l2}[c][Bl][\szzz]{$\frac{\alpha_t\|\vec{v}_t\|_2}{\sqrt{M}}$}
         \psfrag{lam}[l][Bl][\szzz]{$\lambda_t$}
         \psfrag{Ah}[c][Bl][\sz]{$\vec{B}_t$}
         \psfrag{A}[c][Bl][\sz]{$\vec{A}_t$}
         \psfrag{eta}[c][Bl][\szzz]{$\shrinkage{st}\left(\vec{r}_t;\lambda_t\right)$}
         \psfrag{b}[l][Bl][\szzz]{$b_{t+1}$}
        }

\subsection{Parameterizing LAMP-\texorpdfstring{$\ell_1$}{L1}} \label{sec:lampparam}

It is important to realize that LAMP-$\ell_1$ implements a \emph{generalization} of the AMP-$\ell_1$ algorithm \eqref{amp}, wherein the matrices $(\vec{A},\vec{A}\tran)$ manifest as $(\vec{A}_t,\vec{B}_t)$ at iteration $t$.
In other words, the AMP algorithm enforces
$\vec{B}_t=\vec{A}_t\tran$ and 
$\vec{A}_t=\vec{A}_0~\forall t$,
whereas the LAMP-$\ell_1$ network does not.
An important question is whether this generalization preserves the independent-Gaussian nature \eqref{lsl} of the shrinkage input error---the most important feature of AMP.
We will show, numerically, that the desired behavior does seem to occur when 
\begin{align}
\label{eq:simplify}
\vec{A}_t &= \beta_t\vec{A}  
\end{align}
with $\beta_t>0$, at least when $\vec{A}$ is i.i.d.\ Gaussian.

Note that, in \eqref{simplify}, ``$\vec{A}$'' refers to the true measurement matrix from \eqref{y}.  
If $\vec{A}$ was unknown, we could instead use an estimate of $\vec{A}$ computed from the training data, as described in \secref{lamplearn}.
But, in many applications of the sparse linear inverse problem, $\vec{A}$ is known.
Furthermore, if matrix-vector multiplication with $\vec{A}$ was known to have a fast implementation (e.g., FFT), then it could be exploited in \eqref{simplify}.

In \appref{redefine}, we show that, under the parameterization \eqref{simplify} and some redefinitions of variables, the $t^{th}$ layer of the LAMP-$\ell_1$ network can be summarized as
\begin{subequations} \label{eq:lamp}
\begin{align}
\hvec{x}_{t+1}
&= \beta_t \shrinkage{st} \left( \hvec{x}_t + \vec{B}_t\vec{v}_t; \tfrac{\alpha_t}{\sqrt{M}} \|\vec{v}_t\|_2 \right)
\label{eq:lamp2} \\
\vec{v}_{t+1}
&= \vec{y} - \vec{A}\hvec{x}_{t+1} + \tfrac{\beta_{t}}{M}\|\hvec{x}_{t+1}\|_0 \vec{v}_{t}
\label{eq:lamp1} ,
\end{align}
\end{subequations}
with first-layer inputs $\hvec{x}_0=\vec{0}$ and $\vec{v}_0=\vec{y}$.
The LAMP-$\ell_1$ parameters are then
$\vec{\Theta} = \big\{\vec{B}, \{\alpha_t, \beta_t \}_{t=0}^{T-1}\big\}$ in the tied case, or $\vec{\Theta} = \{\vec{B}_t, \alpha_t, \beta_t \}_{t=0}^{T-1}$ in the untied case.

\begin{figure}[t]
	\newcommand{\wid}{0.3\columnwidth}
	\newcommand{\sz}{0.6}
	\newcommand{\szz}{0.5}
	\begin{tabular}{@{}ccc@{}}
        \psfrag{ISTA}[b][B][\sz]{\sf (a) ISTA}
        \psfrag{Standard Normal Quantiles}[t][t][\szz]{\sf Standard Normal Quantiles}
        \psfrag{Quantiles of Input Sample}[b][b][\szz]{\sf Quantiles of Input Sample}
	\includegraphics[width=\wid]{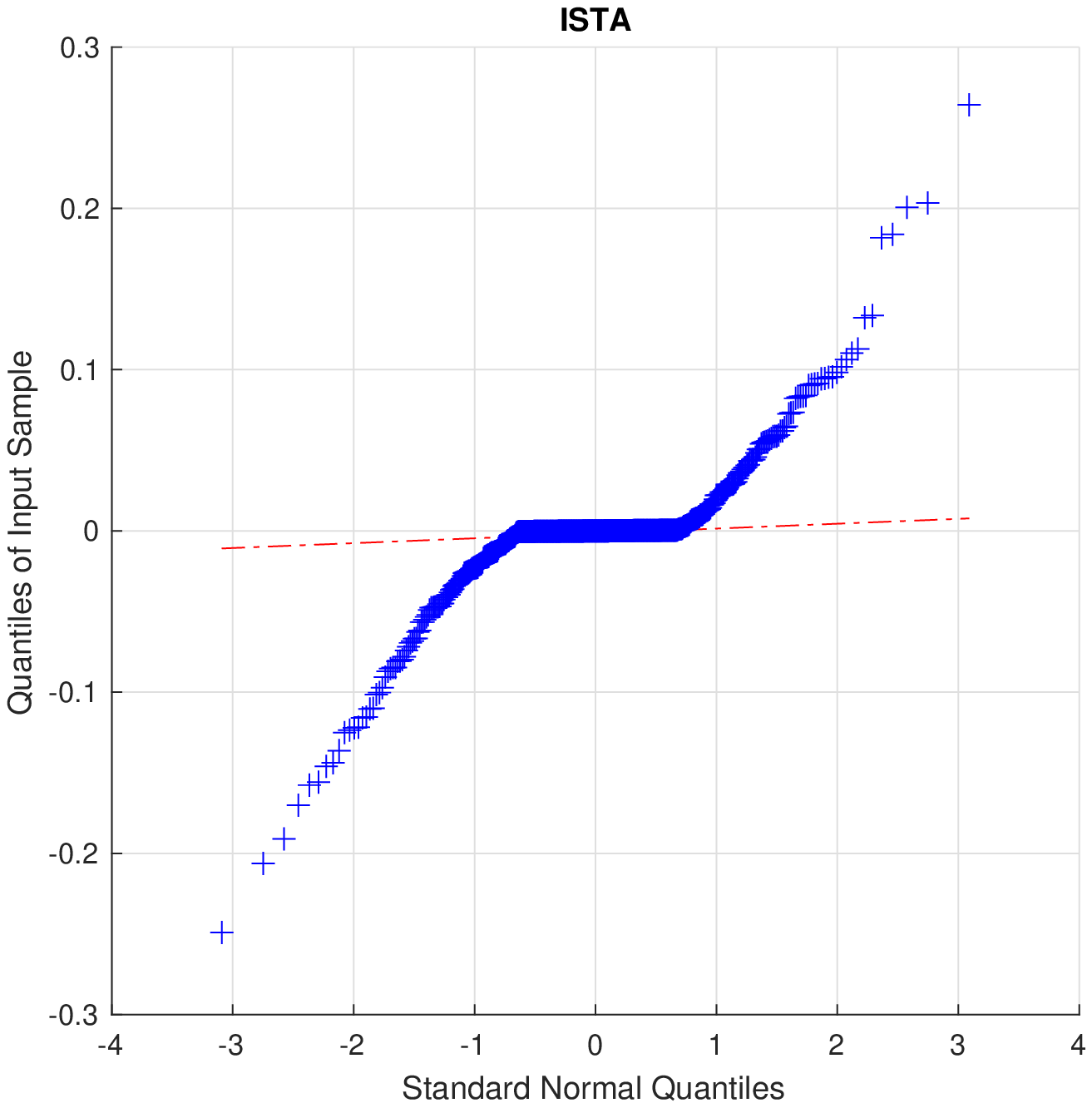}&
        \psfrag{AMP}[b][B][\sz]{\sf (b) AMP-l1}
        \psfrag{Standard Normal Quantiles}[t][t][\szz]{\sf Standard Normal Quantiles}
        \psfrag{Quantiles of Input Sample}[b][b][\szz]{\sf Quantiles of Input Sample}
	\includegraphics[width=\wid]{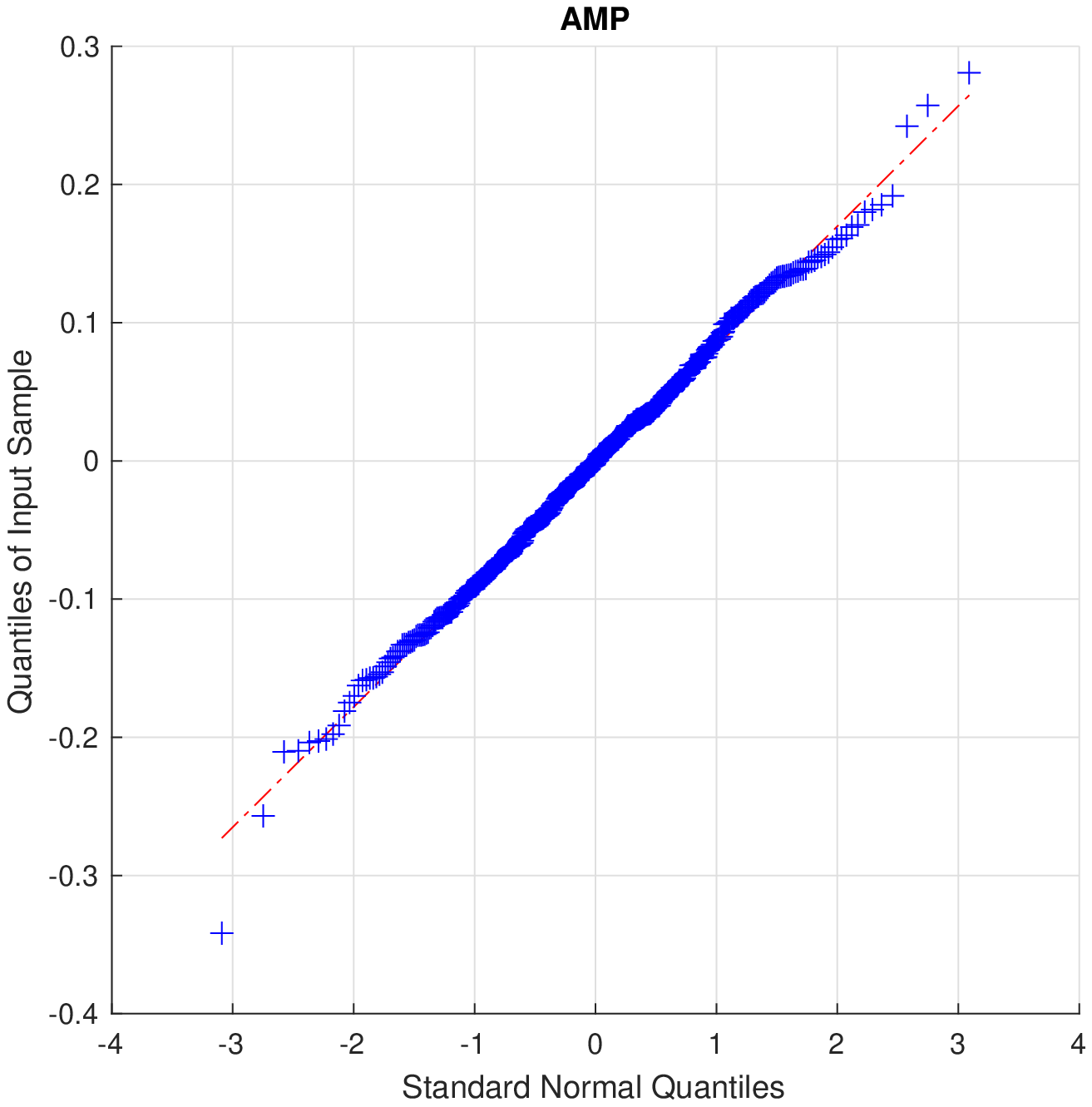}&
        \psfrag{LAMP}[b][B][\sz]{\sf (c) LAMP-l1}
        \psfrag{Standard Normal Quantiles}[t][t][\szz]{\sf Standard Normal Quantiles}
        \psfrag{Quantiles of Input Sample}[b][b][\szz]{\sf Quantiles of Input Sample}
	\includegraphics[width=\wid]{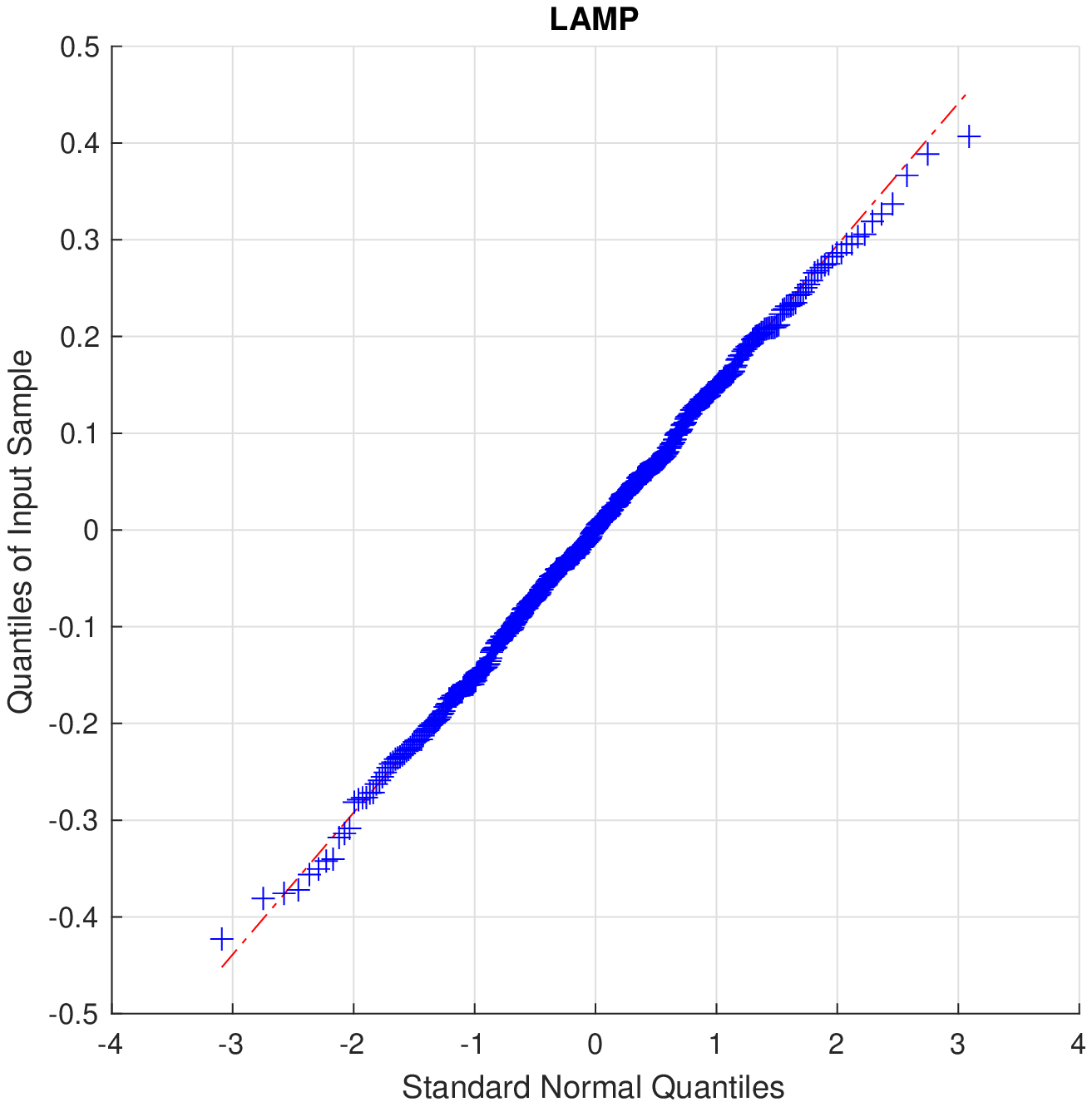}\\
	\end{tabular}
	\caption{QQ plots of the shrinkage input error evaluated at the first iteration/layer $t$ for which NMSE$(\hvec{x}_t)\!<\!-15$~dB 
        (i.e., $t=1478$ for ISTA, $t=6$ for AMP-$\ell_1$, and $t=3$ for untied LAMP-$\ell_1$.)
        The plots show that ISTA's error is heavy tailed while AMP-$\ell_1$'s and LAMP-$\ell_1$'s errors are Gaussian due to Onsager correction.}
	\label{fig:qqplots}
\end{figure}

\Figref{qqplots}(c) shows a quantile-quantile (QQ) plot for the error in the input to untied-LAMP's shrinkage function, 
$(\hvec{x}_t + \vec{B}_t\vec{v}_t)-\vec{x}\true$, 
at a middle layer $t$, using the data from \figref{ista_fista_amp_vamp_Giid_k15}(a).
Also shown are the shrinkage inputs for ISTA and AMP.
The figure shows that the quantiles of AMP-$\ell_1$ and LAMP-$\ell_1$ fall on the dashed diagonal line, confirming that they are Gaussian distributed.
In contrast, the quantiles of ISTA are heavy-tailed.

\subsection{Learning the LAMP-\texorpdfstring{$\ell_1$}{L1} Parameters} \label{sec:lamplearn}

For the ``tied'' case of LAMP-$\ell_1$, we aim to learn the parameters  
$\vec{\Theta}\tied_{T-1} \defn \big\{\vec{B}, \{\alpha_t, \beta_t \}_{t=0}^{T-1}\big\}$
that minimize the MSE on the training data, i.e., \eqref{loss}.
In a first attempt to do this, we tried the standard back-propagation approach, where $\vec{\Theta}\tied_{T-1}$ were jointly optimized from the initialization 
$\vec{B}=\vec{A}\tran,\alpha_0=1,\beta_0=1$, 
but we found that the parameters converged to a bad local minimum.
We conjecture that this failure was a result of overfitting, since $\vec{B}$ had many free parameters in our experiments: $125\,000$, since $\vec{B}\in\Real^{500\times 250}$.
Thus we propose a hybrid of ``layer-wise'' and ``global'' optimization that appears to avoid this problem.

Roughly speaking, our approach is to
learn $\vec{\Theta}\tied_{0}$, 
then $\vec{\Theta}\tied_{1}$, 
and so on, 
until $\vec{\Theta}\tied_{T-1}$. 
Recall that $\vec{\Theta}\tied_t$ are not the parameters of layer $t$ but the parameters of all layers \emph{up to and including} layer $t$.
The details of our approach are specified in \algref{tune_tied}.
There, \lineref{layerwise} performs layer-wise learning (of layer $t$) and \lineref{global} performs global learning (of all layers up to and including $t$).
Note that, in \lineref{Theta0}, we do not learn the parameter $\beta_0$ but instead leave it at its initial value.
The reason is that the triple $\{\vec{B},\alpha_0,\beta_0\}$ is over-parameterized, in that $\{\mu\vec{B},\mu\alpha_0,\beta_0/\mu\}$ 
gives the same layer-$0$ output $\hvec{x}_0$ for any $\mu>0$, due to the property $\shrinkage{st}(\vec{r};\lambda)=\shrinkage{st}(\mu\vec{r};\mu\lambda)/\mu$ of the soft-thresholding function.
To avoid this over-parameterization, we fix the value of $\beta_0$.

\begin{algorithm}
    \caption{Tied LAMP-$\ell_1$ parameter learning}
    \begin{algorithmic}[1]
        \STATE{Initialize $\vec{B}=\vec{A}\tran,\alpha_0=1,\beta_0=1$}
        \STATE{Learn $\vec{\Theta}\tied_0 = \{\vec{B},\alpha_0\}$} \label{line:Theta0}
        \FOR{$t=1$ \TO $T-1$}
        \STATE{Initialize $\alpha_t=\alpha_{t-1},\beta_t=\beta_{t-1}$}
        \STATE{Learn $\{\alpha_t,\beta_t\}$ with fixed $\vec{\Theta}\tied_{t-1}$} \label{line:layerwise} 
        \STATE{Re-learn $\vec{\Theta}\tied_t = \big\{\vec{B},\{\alpha_i,\beta_i\}_{i=1}^t,\alpha_0\big\}$} \label{line:global} 
        \ENDFOR
        \STATE{Return $\vec{\Theta}\tied_{T-1}$ }
    \end{algorithmic}
    \label{alg:tune_tied}
\end{algorithm}

For the untied case of LAMP-$\ell_1$, we aim to learn the parameters
$\vec{\Theta}\untied_{T-1} = \{\vec{B}_t, \alpha_t, \beta_t \}_{t=0}^{T-1}$.
Here we found that extra care was needed to avoid bad local minima.
To this end, we implemented a bootstrapping method based on the following rationale:
a network that can choose a different $\vec{B}_t$ for each layer $t$ should perform at least as well as one that is constrained to use the same $\vec{B}$ for all layers $t$.
In particular, our bootstrapping method checks performance against tied LAMP-$\ell_1$ at each layer $t$ and reinitializes using the tied parameters when appropriate.
The details are given in \algref{tune_untied}.

\begin{algorithm}
    \caption{Untied LAMP-$\ell_1$ parameter learning}
    \begin{algorithmic}[1]
        \STATE{Compute $\{\Vec{\Theta}\tied_t\}_{t=1}^{T-1}$ using \algref{tune_tied}} 
        \STATE{Initialize $\vec{B}_0=\vec{A}\tran,\alpha_0=1,\beta_0=1$}
        \STATE{Learn $\Vec{\Theta}\untied_0 = \{\vec{B}_0,\alpha_0\}$} 
        \FOR{$t=1$ \TO $T-1$}
        \STATE{Initialize $\vec{B}_t=\vec{B}_{t-1},\alpha_t=\alpha_{t-1},\beta_t=\beta_{t-1}$}
        \STATE{Learn $\{\vec{B}_t,\alpha_t,\beta_t\}$ with fixed $\Vec{\Theta}\untied_{t-1}$ } 
        \STATE{Set $\vec{\Theta}\untied_t = \{\vec{B}_i,\alpha_i,\beta_i\}_{i=0}^t \setminus \beta_0$}  
        \IF{ $\Vec{\Theta}\tied_t$ performs better than $\vec{\Theta}\untied_t$}
            \STATE{Replace $\vec{\Theta}\untied_t$ with $\Vec{\Theta}\tied_t$ (setting $\vec{B}_i=\vec{B}~\forall i$)}
        \ENDIF
        \STATE{Re-learn $\vec{\Theta}\untied_t$}  
        \ENDFOR
        \STATE{Return $\vec{\Theta}\untied_{T-1}$ }
    \end{algorithmic}
    \label{alg:tune_untied}
\end{algorithm}

As described in \secref{lampnet}, our LAMP-$\ell_1$ parameterization \eqref{simplify} assumes that $\vec{A}$ is known.
If $\vec{A}$ is unknown, it could be estimated using a least-squares (LS) fit\footnote{For the least-squares learning of $\vec{A}$, one could either use the one-shot approach $\hvec{A}=\vec{YX}^+$ where $\vec{Y}=[\vec{y}\of{1},...,\vec{y}\of{D}]$ and $\vec{X}=[\vec{x}\of{1},...,\vec{x}\of{D}]$ and $\vec{X}^+$ is the pseudo-inverse of $\vec{X}$, or one could use back-propagation to minimize the loss $\sum_{d=1}^D \|\vec{y}\of{d}-\vec{A}\vec{x}\of{d}\|_2^2$.}
to the training data and further optimized along with the parameters $\vec{\Theta}\tied_{T-1}$ or $\vec{\Theta}\untied_{T-1}$ to minimize the loss $\mc{L}_T$ from \eqref{loss}.
Empirically, we find (in experiments not detailed here) that there is essentially no difference between the final test MSEs of LAMP-$\ell_1$ networks trained with known $\vec{A}$ or LS-estimated $\vec{A}$.

\subsection{Discussion} \label{sec:discussion1}

In this section, we proposed a LAMP network whose nonlinear stages were constrained to the soft-thresholding shrinkage $\shrinkage{st}(\cdot)$ from \eqref{soft_thresh}. 
Under this constraint, the resulting LAMP-$\ell_1$ network differs from LISTA only in the presence of Onsager correction, allowing us to study the effect of Onsager correction in deep networks.
The numerical experiments in \secref{numerical1} show that, as expected, the LAMP-$\ell_1$ network outperforms the LISTA network at every layer for the numerical data used to create \figref{ista_fista_amp_vamp_Giid_k15}.

\section{Learned AMP} \label{sec:lampgen}

We now consider the use of generic shrinkage functions $\shrinkage{}(\cdot)$ within LAMP with the goal of improving its performance over that of LAMP-$\ell_1$.
In particular, we aim to learn the jointly MSE-optimal shrinkage functions and linear transforms across all layers of the LAMP network.
To make this optimization tractable, we consider several \emph{families} of shrinkage functions, where each family is parameterized by a finite-dimensional vector $\vec{\theta}_t$ at layer $t$. 
We then use back-propagation to learn the jointly MSE-optimal values of $\{\vec{\theta}_t\}_{t=0}^{T-1}$ and the linear-transform parameters.

\subsection{The LAMP Network} \label{sec:lampgennet}

For LAMP, we unfold the generic AMP algorithm \eqref{ampgen} into a network.
As with AMP-$\ell_1$, we relax the linear transform pair $(\vec{A},\vec{A}\tran)$ to the layer-dependent learnable pair $(\vec{A}_t,\vec{B}_t)$, and then place the restrictions on $\vec{A}_t$ 
to facilitate Onsager correction.
With AMP-$\ell_1$, the restrictions came in the form of \eqref{simplify}, where $\beta_t$ and $\vec{B}_t$ emerged as the tunable parameters.
It was then shown, in \appref{redefine}, that $\beta_t$ acted to scale the output of the soft-thresholding function.
Since the shrinkage functions that we use in this section will have their own scaling mechanisms, it now suffices to use \eqref{simplify} with $\beta_t=1$. 
Under this parameterization, the $t$th layer of (general) LAMP becomes
\begin{subequations} \label{eq:lampgen}
\begin{align}
\hvec{x}_{t+1}
&= \shrinkage{} \left( \hvec{x}_t + \vec{B}_t\vec{v}_t; \sigma_t, \vec{\theta}_t \right) \\
\vec{v}_{t+1}
&= \vec{y} - \vec{A}\hvec{x}_{t+1} + b_{t+1} \vec{v}_{t} ,
\end{align}
\end{subequations}
with learnable parameters $\vec{B}_t$ and $\vec{\theta}_t$.
See \figref{lampGen_1layer} for an illustration.

\putFrag{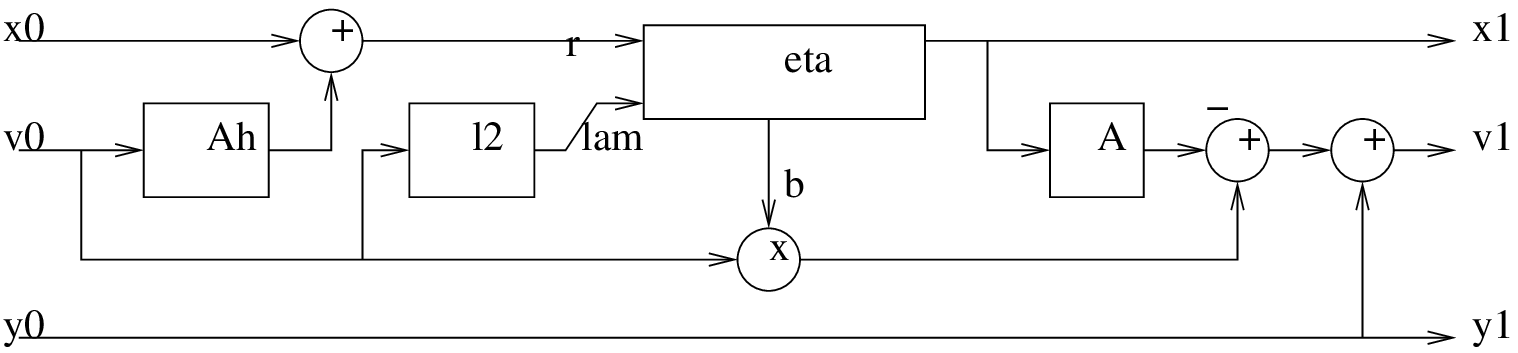}
        {The $t$th layer of the (general) LAMP network, with learnable parameters $\vec{B}_t$ and $\vec{\theta}_t$.}
        {3.15}
        {\newcommand{\sz}{1.0}
         \newcommand{\szz}{0.9}
         \newcommand{\szzz}{0.75}
         \psfrag{+}[c][Bl][\sz]{$+$}
         \psfrag{-}[r][Bl][\szz]{$-$}
         \psfrag{x}[c][Bl][\sz]{$\times$}
         \psfrag{x0}[r][Bl][\sz]{$\hvec{x}_t$}
         \psfrag{x1}[l][Bl][\sz]{$\hvec{x}_{t\!+\!1}$}
         \psfrag{v0}[r][Bl][\sz]{$\vec{v}_t$}
         \psfrag{v1}[l][Bl][\sz]{$\vec{v}_{t\!+\!1}$}
         \psfrag{y0}[r][Bl][\sz]{$\vec{y}$}
         \psfrag{y1}[l][Bl][\sz]{$\vec{y}$}
         \psfrag{r}[t][Bl][\szzz]{$\vec{r}_t$}
         \psfrag{l2}[c][Bl][\szzz]{$\frac{\|\vec{v}_t\|_2}{\sqrt{M}}$}
         \psfrag{lam}[l][Bl][\szzz]{$\sigma_t$}
         \psfrag{Ah}[c][Bl][\szz]{$\vec{B}_t$}
         \psfrag{A}[c][Bl][\sz]{$\vec{A}$}
         \psfrag{eta}[c][Bl][\szzz]{$\shrinkage{}\left(\vec{r}_t;\sigma_t,\vec{\theta}_t\right)$}
         \psfrag{b}[l][Bl][\szzz]{$b_{t+1}$}
        }

\subsection{Parameterizing the Shrinkage Functions} \label{sec:shrinkage}

In the sequel, we consider families of shrinkage functions 
$\shrinkage{}\left(\vec{r};\sigma,\vec{\theta}\right)$ 
that are both separable and odd symmetric.
By separable, we mean that 
$\left[\shrinkage{}\left(\vec{r};\sigma,\vec{\theta}\right)\right]_j=\eta(r_j;\sigma,\vec{\theta})~\forall j$ for some scalar function $\eta$,
and by odd symmetric we mean that
$\shrinkage{}\left(\vec{r};\sigma,\vec{\theta}\right)= -\shrinkage{}\left(-\vec{r};\sigma,\vec{\theta}\right)$ for all $\vec{r}\in\Real^N$.
Several such shrinkage families are detailed below. 
\smallskip

\subsubsection{Scaled Soft-Threshold} 
We first consider 
\begin{align} \label{eq:sst}
\left[ \shrinkage{sst}\left(\vec{r};\sigma,\vec{\theta}\right) \right]_j
\defn
\theta_1 \sgn(r_j)  \max\{|r_j|- \theta_2\sigma , 0 \} ,
\end{align}
which can be recognized as a scaled version of the soft-threshold operator from \eqref{soft_thresh}.
Note that $\vec{\theta} \in \Real^2$.
It can be readily seen that LAMP-$\ell_1$ from \eqref{lamp} is a special case of LAMP from \eqref{lampgen} for which $\shrinkage{} = \shrinkage{sst}\,$ and $\vec{\theta}_t=[\beta_t,\alpha_t]$. 

\subsubsection{Piecewise Linear}
Next we consider (odd symmetric) piecewise linear functions with five segments:
\begin{align} \label{eq:pwlin}
\lefteqn{
\left[\shrinkage{pwlin}(\vec{r};\sigma,\vec{\theta})\right]_j
}\\
&\defn \begin{cases}
\theta_3 r_j 
& \text{if~} |r_j| \leq \theta_1\sigma \\ 
\begin{array}{@{}l@{}} 
\sgn(r_j) \big[ \theta_4(|r_j|-\theta_1\sigma) \\
\mbox{}\quad + \theta_3\theta_1\sigma\big] 
\end{array}
& \text{if~} \theta_1\sigma < |r_j| \leq \theta_2\sigma \\ 
\begin{array}{@{}l@{}} 
\sgn(r_j) \big[ \theta_5(|r_j|-\theta_2\sigma) \\
\mbox{}\quad + \theta_4(\theta_2-\theta_1)\sigma + \theta_3\theta_1\sigma\big] 
\end{array}
& \text{if~} \theta_2\sigma < |r_j| .
\end{cases}
\nonumber
\end{align}
Here, the shrinkage-family parameters $\vec{\theta} \in \Real^5$ determine 
the abscissae of the four vertices where the line segments meet 
(i.e., $[-\theta_2\sigma,-\theta_1\sigma,\theta_1\sigma,\theta_2\sigma]$) 
and the slopes of the five segments 
(i.e., $[\theta_5,\theta_4,\theta_3,\theta_4,\theta_5]$).
The shrinkage in \eqref{pwlin} can be considered as a generalization of \eqref{sst} from three to five segments with a possibly non-zero slope on the middle segment.
It is inspired by the design from \cite[Eq. (13)-(15)]{GuoDavies:TSP:15} but has a different parameterization and includes a dependence on the estimated noise level $\sigma$.

\subsubsection{Exponential}
We now consider the exponential shrinkage family
\begin{align} \label{eq:exp}
\left[ \shrinkage{exp}\left(\vec{r};\sigma,\vec{\theta}\right) \right]_j
&\defn \theta_2 r_j + \theta_3 r_j \exp\left(-\frac {r_j^2} {2 \theta_1^2 \sigma^2 } \right) .
\end{align}
The parameters $\vec{\theta} \in \Real^3$ control
the asymptotic slope (i.e., $\theta_2$),
the slope at the origin (i.e., $\theta_2+\theta_3$), 
and the rate of transition between those two slopes (where larger $\theta_1$ gives a slower transition).
The shrinkage in \eqref{exp} is inspired by the design from \cite[Eq. (19)-(20)]{GuoDavies:TSP:15} but includes a dependence on the estimated noise level $\sigma$. 

\subsubsection{Spline}
Next we consider the spline shrinkage family
\begin{align} \label{eq:spline}
\left[ \shrinkage{spline}\left(\vec{r};\sigma,\vec{\theta}\right) \right]_j
&\defn \theta_2 r_j + \theta_3 r_j \beta \left( \frac {r_j} {\theta_1 \sigma}  \right) ,
\end{align}
where $\beta$ is the cubic B-spline \cite{Unser:SPM:99}
\begin{equation}
\beta(z) \defn 
\begin{cases}
\frac{2}{3}-|z|^2 + \frac{|z|^3}{2}  & \text{if~} 0 \leq |z| \leq 1\\
\frac{1}{6}(2 - |z|)^3 & \text{if~} 1 \leq |z| \leq 2 \\
0 & \text{if~} 2 \leq |z| .
\end{cases}
\end{equation}
Similar to \eqref{exp},
the parameters $\vec{\theta} \in \Real^3$ in \eqref{spline} control
the asymptotic slope (i.e., $\theta_2$),
the slope at the origin (i.e., $\theta_2+\tfrac{2}{3}\theta_3$), 
and the rate of transition between those two slopes (where larger $\theta_1$ gives a slower transition).
The shrinkage in \eqref{spline} is inspired by that used in \cite{Kamilov:SPL:16}, but is parameterized differently.
The shrinkage in \cite{Kamilov:SPL:16} was constructed using $8000$ shifts of $\beta(z)$ spread uniformly over the dynamic range of the signal, each scaled by an adjustable weight.
By contrast, the shrinkage in \eqref{spline} has only three adjustable parameters but includes a dependence on the noise level $\sigma$.
Furthermore, \cite{Kamilov:SPL:16} used identical shrinkage parameters at all layers of the ISTA network, whereas we allow the shrinkage parameters $\vec{\theta}$ to vary across the layers of the LAMP network.

\subsubsection{Bernoulli-Gaussian}

Finally, we consider shrinkage functions that correspond to MSE-optimal denoisers under zero-mean Bernoulli-Gaussian (BG) priors.
That is, $\hat{x}=\E\{x|r\}$, where $x$ has the BG prior
\begin{align}
p(x;\gamma,\phi) 
&= (1-\gamma)\delta(x) + \gamma \mc{N}(x;0,\phi) 
\end{align}
(with $\gamma\in(0,1)$ and $\phi>0$)
and $r$ is an AWGN-corrupted measurement of $x$:
\begin{align}
r=x+e \text{~~for~~} e\sim\mc{N}(0,\sigma^2)
\label{eq:awgn} .
\end{align}
The MSE-optimal denoiser is then (see, e.g., \cite{Vila:TSP:13}) 
\begin{align}
\hat{x}
= \frac{r}{\Big(1+\frac{\sigma^2}{\phi}\Big)
        \big(1+\frac{1-\gamma}{\gamma}
                \frac{\mc{N}(r;0,\sigma^2)}{\mc{N}(r;0,\sigma^2+\phi)}\Big)} 
\label{eq:BGdenoise} .
\end{align}

To turn \eqref{BGdenoise} into a learnable shrinkage function, we set
$\theta_1=\phi$ and $\theta_2=\log\frac{1-\gamma}{\gamma}$ and then simplify, giving 
\begin{align}
\lefteqn{
\left[ \shrinkage{bg}\left(\vec{r};\sigma,\vec{\theta}\right) \right]_j
} \label{eq:bgest} \\
&= \frac{r_j}{\Big(1+\frac{\sigma^2}{\theta_1}\Big)
        \Big(1+
        \sqrt{1+\frac{\theta_1}{\sigma^2}}
        \exp\left[\theta_2-\frac{r_j^2}{2(\sigma^2+\sigma^4/\theta_1)}\right]
        \Big)} 
\nonumber .
\end{align}

\subsection{Learning the LAMP Parameters} \label{sec:lampgenlearn}

As with LAMP-$\ell_1$, we consider two cases of LAMP: the ``tied'' case, where the same linear transform is used at all layers of the network, and the ``untied'' case where a different linear transform is allowed in each layer.
Thus, the parameters for the tied LAMP are $\big\{\vec{B},\{\vec{\theta}_t\}_{t=0}^{T-1}\big\}$ and those for untied LAMP are $\{\vec{B}_t,\vec{\theta}_t\}_{t=0}^{T-1}$.
The LAMP parameters are then learned using the method described in \secref{lamplearn}, now with $\{\alpha_t,\beta_t\}$ replaced by $\vec{\theta}_t$.

\subsection{Discussion} \label{sec:discussion2}

In this section, we constructed a ``LAMP'' deep network by unfolding the AMP algorithm from \cite{Donoho:PNAS:09}, parameterizing its linear and nonlinear stages in novel ways, and learning the parameters using a hybrid of layer-wise and global optimization. 
The numerical experiments in \secref{numerical} suggest that LAMP performs quite well with i.i.d.\ Gaussian $\vec{A}$.
For example, after $10$ layers, untied LAMP's NMSE is $0.5$~dB from the support-oracle bound and as much as $16$~dB better than that of the (tied) LISTA approach from \cite{Gregor:ICML:10}. 

For non-i.i.d.-Gaussian $\vec{A}$, and especially ill-conditioned $\vec{A}$, however, the performance of LAMP suffers.
Also, it is not clear how to interpret the parameters learned by LAMP, even in the case of i.i.d.\ Gaussian $\vec{A}$. 
Both problems stem from the fact that LAMP can be viewed as a generalization of AMP that uses the matrices 
$(\beta_t\vec{A},\vec{B}_t)$ in place of $(\vec{A},\vec{A}\tran)$ at the $t$ iteration.
We aim to resolve these issues using the method presented in the next section.

\section{Learned Vector-AMP} \label{sec:lvamp}

As described in \secref{algorithms}, the behavior of AMP is well understood when $\vec{A}$ is i.i.d.\ sub-Gaussian, but even small deviations from this model can lead AMP to diverge or at least behave in ways that are not well understood.
Very recently, however, the VAMP algorithm has been proposed as a partial solution to this problem.
That is, VAMP enjoys the same benefits of AMP but works with a much larger class of matrices $\vec{A}$: those that are right-rotationally invariant.
Perhaps, by building a deep network around the VAMP algorithm, we can circumvent the problems with LAMP that arise with non-i.i.d.-Gaussian matrices. 

\subsection{The LVAMP Network} \label{sec:lvampnet}

We propose to unfold the VAMP algorithm into a network and learn the MSE-optimal values of its parameters.
The $t$th layer of the learned VAMP (LVAMP) network is illustrated in \figref{lvamp_1layer}.
Essentially it consists of four operations: 
1) LMMSE estimation, 2) decoupling, 3) shrinkage, and 4) decoupling,
where the two decoupling stages are identical.

With an i.i.d.\ signal, the LMMSE estimator takes the form \eqref{lmmse2}. 
Plugging the SVD \eqref{svd} into \eqref{lmmse2} yields \eqref{lmmse}. 
Thus, since VAMP assumes an i.i.d.\ signal, its LMMSE stage is parameterized by $\tvec{\theta}=\{\vec{U},\vec{s},\vec{V},\sigma_w^2\}$ for all iterations $t$ (recall \eqref{thetatil}).
For generality, we allow the LVAMP to vary these parameters with the layer $t$, giving $\tvec{\theta}_t=\{\vec{U}_t,\vec{s}_t,\vec{V}_t,\sigma_{wt}^2\}$.

With non-i.i.d.\ (e.g., correlated) signals, the LMMSE estimator also depends on the signal covariance matrix, which may not be explicitly known.
In this case, it makes more sense to parameterize LVAMP's layer-$t$ LMMSE stage as
\begin{align}
\tvec{\eta}\big(\tvec{r}_t;\tilde{\sigma}_t,\tvec{\theta}_t\big)
&= \vec{G}_t \tvec{r}_t + \vec{H}_t \vec{y}
\label{eq:lmmse3} 
\end{align}
with unconstrained $\vec{G}_t\in\Real^{N\times N}$ and $\vec{H}_t\in\Real^{N\times M}$,
in which case $\tvec{\theta}_t=\{\vec{G}_t,\vec{H}_t\}$.
In either case,
the nonlinear stage is characterized by the shrinkage parameters $\vec{\theta}_t$, whose format depends on which shrinkage family is being used.

\subsection{Learning the LVAMP Parameters} \label{sec:lvamplearn}

As before, one can imagine ``tied'' and ``untied'' network parameterizations.
In the tied case, the network parameters would be
$\big\{\tvec{\theta},\{\vec{\theta}_t\}_{t=1}^T\big\}$, 
while in the untied case, they would be
$\{\tvec{\theta}_t,\vec{\theta}_t\}_{t=1}^T$.
But note that, with the SVD parameterization of $\tvec{\eta}(\cdot)$, even tied parameters $\tvec{\theta}$ yield an LMMSE estimator \eqref{lmmse} that varies with the layer $t$ due to its dependence on $\tilde{\sigma}_t$. 

To learn the LVAMP parameters, we propose to use \algref{tune_tied} for the tied case and \algref{tune_untied} for the untied case (with $\tvec{\theta}_t$ replacing $\vec{B}_t$ and with $\vec{\theta}_t$ replacing $\{\alpha_t,\beta_t\}$).
When $\vec{A}$ is known, we suggest to initialize $\{\vec{U},\vec{s},\vec{V}\}$ at the SVD values from \eqref{svd}.
When $\vec{A}$ is unknown, we suggest to initialize with an SVD of the least-squares estimate of $\vec{A}$ from the training data, as discussed in \secref{lamplearn}.
Finally, we suggest to initialize $\sigma_w^2$ at the average value of $M^{-1}\|\vec{y}\|^2$ across the training data.

\putFrag{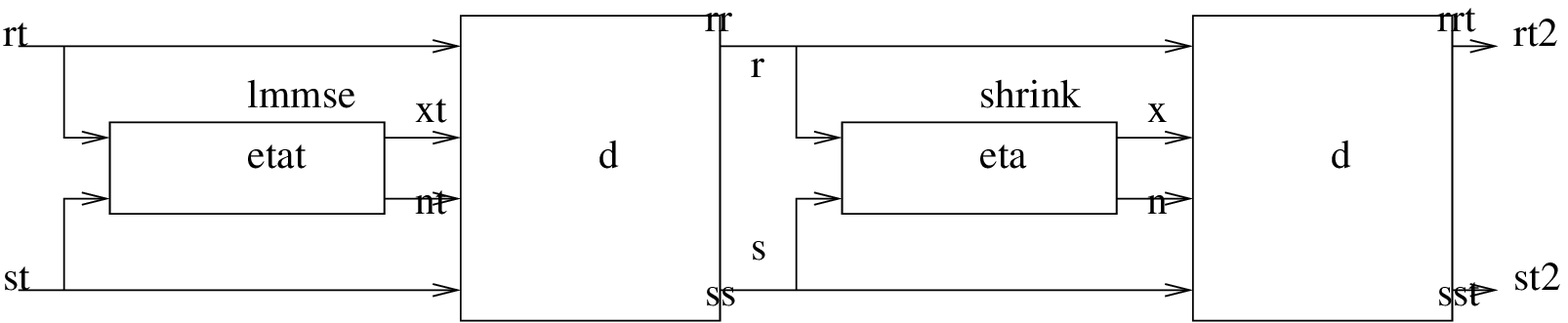}
        {The $t$th layer of the LVAMP network, with learnable LMMSE parameters $\tvec{\theta}_t$ and learnable shrinkage parameters $\vec{\theta}_t$.}
        {3.15}
        {\newcommand{\sz}{1.0}
         \newcommand{\szz}{0.82}
         \newcommand{\szzz}{0.75}
         \newcommand{\szzzz}{0.65}
         \psfrag{x}[b][Bl][\szz]{~$\hvec{x}_t$}
         \psfrag{xt}[b][Bl][\szz]{~$\tvec{x}_t$}
         \psfrag{n}[t][Bl][\szz]{~$\nu_t$}
         \psfrag{nt}[t][Bl][\szz]{~$\tilde{\nu}_t$}
         \psfrag{r}[][Bl][\szz]{$\vec{r}_t$}
         \psfrag{rt}[r][Bl][\sz]{$\tvec{r}_t$}
         \psfrag{rt2}[l][Bl][\sz]{$\tvec{r}_{t+1}$}
         \psfrag{s}[][Bl][\szz]{$\sigma_t$}
         \psfrag{st}[r][Bl][\sz]{$\tilde{\sigma}_t$}
         \psfrag{st2}[l][Bl][\sz]{$\tilde{\sigma}_{t+1}$}
         \psfrag{etat}[c][Bl][\szzz]{$\tvec{\eta}\big(\tvec{r}_t;\tilde{\sigma}_t,\tvec{\theta}_t\big)$}
         \psfrag{eta}[c][Bl][\szzz]{$\vec{\eta}\big(\vec{r}_t;\sigma_t,\vec{\theta}_t\big)$}
         \psfrag{rr}[tr][Bl][\szz]{$\frac{\tvec{x}_t-\tilde{\nu}_t\tvec{r}_t}{1-\tilde{\nu}_t}\!\!=\!$}
         \psfrag{rrt}[tr][Bl][\szz]{$\frac{\hvec{x}_t-\nu_t\vec{r}_t}{1-\nu_t}\!\!=\!$}
         \psfrag{ss}[br][Bl][\szzz]{$\tilde{\sigma}_t\frac{\sqrt{\tilde{\nu}_t}}{\sqrt{1-\tilde{\nu}_t}}\!\!=\!$}
         \psfrag{sst}[br][Bl][\szzz]{$\sigma_t\frac{\sqrt{\nu_t}}{\sqrt{1-\nu_t}}\!\!=\!$}
         \psfrag{lmmse}[B][Bl][\szzzz]{\sf LMMSE}
         \psfrag{shrink}[B][Bl][\szzzz]{\sf shrinkage}
         \psfrag{d}[][Bl][\szzzz]{\sf decouple}
        }

\subsection{Discussion} \label{sec:discussion3}

The numerical results in \secref{numerical} show that, with i.i.d. signals and the SVD parameterization of $\tvec{\theta}$, the tied and untied versions of LVAMP perform near-identically.
Furthermore they show that, as conjectured, the LVAMP network is much more robust to the matrix $\vec{A}$ than the LAMP network.
And even for i.i.d.\ Gaussian $\vec{A}$, LVAMP converges a bit faster than LAMP to a near-oracle MSE level.

Perhaps even more interesting is the finding that, with i.i.d. signals, the parameter values learned by the LVAMP network are \emph{essentially identical} to the ones prescribed by the matched VAMP algorithm.
Thus, the interpretability of the VAMP algorithm (i.e., the fact that it alternates between linear MMSE vector estimation and non-linear MMSE scalar estimation) translates directly to the LVAMP network.
These and other findings will be discussed in more detail in \secref{numerical4}.

\subsection{Summary of Computational and Memory Complexity}

We now outline the complexity and memory costs of the $T$-layer LISTA, LAMP, and LVAMP networks, assuming that $M\ll N$ and $|\vec{\theta}|\ll M^2$, where $|\vec{\theta}|$ denotes the number of shrinkage parameters in $\vec{\theta}$.
See \tabref{cost} for a summary.

Untied LISTA learns $\vec{B}\in\Real^{N\times M}$, $\vec{S}_t\in\Real^{N\times N}$, and $\vec{\theta}_t$ for $t=1\dots T$ and does one matrix-vector multiply with $\vec{S}_t$ in the $t$ layer.
Thus, if $M\ll N$, its computational and memory complexities are $\approx TN^2$ over $T$ stages.
Tied LISTA is similar except that there is only one $\vec{S}$ to learn, reducing its memory complexity to $N^2$.
 
Untied LAMP learns $\vec{B}_t\in\Real^{N\times M}$ and $\vec{\theta}_t$ for $t=1\dots T$ and does one matrix-vector multiply with $\vec{B}_t$ and with $\vec{A}$ in the $t$th layer.
Thus, its computational complexity is $\approx 2TNM$ and its memory complexity is $\approx TMN$ over $T$ stages.
Tied LAMP is similar except that there is only one $\vec{B}$ to learn, reducing its memory complexity to $MN$.
 
For LVAMP with i.i.d.\ signals and SVD-parameterized $\tvec{\theta}$, we saw that untied and tied versions performed nearly identically.
Furthermore, their learned parameters coincided with the ones prescribed by the matched VAMP algorithm.
Thus, there is no need for LVAMP to learn and store the $\vec{U},\vec{s},\vec{V}$ quantities in $\tvec{\theta}$, since they are known.
LVAMP needs to learn and store only $\sigma_w^2$ and the shrinkage parameters $\{\vec{\theta}_t\}_{t=1}^T$, for a total memory complexity of $\approx T|\vec{\theta}|$.
Meanwhile, each layer does a matrix-vector multiply with $\vec{V}$ and $\vec{V}\tran$, since $\vec{Uy}$ can be computed in advance.
Thus the computational complexity over $T$ layers is $\approx 2TNM$.
With the $(\vec{G}_t,\vec{H}_t)$-parameterized $\tvec{\theta}$, the computational and memory complexities would both be $\approx TN^2$, as with untied LISTA.

Finally, we note that the computational complexities of LAMP and LVAMP decrease when $\vec{A}$ and $\vec{V}$ (or $\vec{G}_t,\vec{H}_t$) have fast implementations (e.g., FFT).

\putTable{cost}{Approximate computational complexity (per vector input $\vec{y}$) and memory complexity for $T$-layer networks.}
{\begin{tabular}{c|c|c|c|c|c}
 & untied & tied  & untied & tied & untied \\
 & LISTA  & LISTA &  LAMP  & LAMP & LVAMP  \\\hline
 \begin{tabular}{@{}c@{}}computational\\ complexity\end{tabular}
 & $TN^2$ & $TN^2$ & $2TNM$ & $2TNM$ & $2TNM$ \\\hline
 \begin{tabular}{@{}c@{}}memory\\ complexity\end{tabular}
 & $TN^2$ & $N^2$ & $TMN$ & $MN$ & $T|\vec{\theta}|$
 \end{tabular}}

\section{Numerical Investigation} \label{sec:numerical}

We now investigate 
the effects of learning, 
Onsager correction,
choice of shrinkage $\vec{\eta}(\cdot)$,
network untying, and
matrix $\vec{A}$
through a sequence of experiments on synthetic data.
The data was constructed in the same way as that for \figref{ista_fista_amp_vamp_Giid_k15}, which we review now for convenience.

Recall the sparse linear inverse problem \eqref{y}.
For both training and test data, we constructed random realizations of BG-distributed sparse $\vec{x}\true$ by drawing its elements i.i.d.\ $\mc{N}(0,1)$ with probability $\gamma=0.1$ and otherwise setting them equal to zero.
Likewise, we generated random noise vectors $\vec{w}$ with i.i.d.\ $\mc{N}(0,\sigma_w^2)$ elements, with $\sigma_w^2$ set to yield an SNR $\E\{\|\vec{Ax}\true\|^2\}/\E\{\|\vec{w}\|^2\}$ of $40$~dB.
We considered two realizations of random $\vec{A}\in\Real^{M\times N}$ with $M=250$ and $N=500$.
The first was i.i.d.\ Gaussian, with elements distributed $\mc{N}(0,M^{-1})$ so that $\|\vec{A}\|_F^2\approx N$ (i.e., the scaling expected by AMP).
The second was constructed to have condition number $\kappa(\vec{A})=15$.
To construct this latter matrix, we started with the i.i.d.\ Gaussian $\vec{A}$ and replaced its singular values $s_i$ by a sequence constructed so that $s_i/s_{i-1}=\rho~\forall i>1$, with $\rho$ and $s_1$ chosen so that $s_1/s_M=15$ and $\|\vec{A}\|_F^2=N$.

We used mini-batches of size $D\!=\!1000$ for training and a single mini-batch of size $1000$ for testing (drawn independent of the training data, but from the same distribution).
The training and testing methods were implemented\footnote{Our Python- and Matlab-based implementation can be downloaded from\newline \url{https://github.com/mborgerding/onsager_deep_learning}} in Python using TensorFlow \cite{Abadi:tensorflow:15} with the Adam optimizer \cite{Kingma:ICLR:15}.

\subsection{Effect of Onsager Correction and Parameter Learning}  
\label{sec:numerical1}

First we study the effect of Onsager correction on deep networks.
We do this by comparing the performance of LAMP-$\ell_1$ and LISTA, which differ only in the use of Onsager correction. 
Simultaneously, we study the effect of parameter learning.
We do this by comparing the performance of LAMP-$\ell_1$ and AMP-$\ell_1$, which differ only in the use of parameter learning.
For LAMP-$\ell_1$, we performed the learning as described in \secref{lamplearn}.
For LISTA, we used the same approach to learn 
``tied'' $\vec{\Theta}=\big\{\vec{B},\vec{S},\{\lambda_t\}_{t=0}^{T-1}\big\}$
and ``untied'' $\vec{\Theta}=\big\{\vec{B},\{\vec{S}_t,\lambda_t\}_{t=0}^{T-1}\big\}$,
with no constraints on $\vec{S}_t$ or $\vec{B}$.

\Figref{lista_vs_lamp_Giid} shows average test-NMSE versus layer $t$ for i.i.d.\ Gaussian $\vec{A}$.
The figure shows tied LAMP-$\ell_1$ significantly outperforming both tied LISTA and AMP-$\ell_1$ at each layer.
For example, to reach NMSE $=-34$~dB,
AMP-$\ell_1$ took $25$ iterations (see also \figref{ista_fista_amp_vamp_Giid_k15}(a)),
tied-LISTA took $15$ layers, and
tied-LAMP-$\ell_1$ took only $7$ layers.

\Figref{lista_vs_lamp_k15} shows the corresponding results for $\vec{A}$ with condition number $\kappa=15$.
For this $\vec{A}$, AMP-$\ell_1$ diverged (see also \figref{ista_fista_amp_vamp_Giid_k15}(b)) but LAMP-$\ell_1$ did not.
Rather, tied LAMP-$\ell_1$ gave roughly the same performance relative to tied LISTA as it did for the i.i.d.\ Gaussian case of $\vec{A}$.

These figures also show that the untied versions of LAMP-$\ell_1$ and LISTA yielded modest improvements over the tied versions for i.i.d.\ Gaussian $\vec{A}$ (i.e., $\leq 2$~dB in \figref{lista_vs_lamp_Giid}) and more significant benefits for $\vec{A}$ with $\kappa=15$ (i.e., $\leq 3$~dB in \figref{lista_vs_lamp_k15}).
However, the untied versions incur a $T$-fold increase in parameter storage and significantly increased training time.
Note that the greatest beneficiary of the untied configuration was LAMP-$\ell_1$ with non-i.i.d.-Gaussian $\vec{A}$.  
We conjecture that the LAMP-$\ell_1$ network somehow used the extra freedom available in the untied case to counteract the non-i.i.d.-Gaussian nature of $\vec{A}$.

\putFrag{lista_vs_lamp_Giid}
	{Test NMSE versus layer (or versus iteration for AMP) under i.i.d.\ Gaussian $\vec{A}$.}
	{\figsize}
	{\newcommand{\sz}{0.53}
	 \newcommand{\szz}{0.7}
         \psfrag{AMP}[l][l][\sz]{\sf \!AMP-l1}
         \psfrag{LISTA}[l][l][\sz]{\sf \!LISTA tied}
         \psfrag{LISTA untied}[l][l][\sz]{\sf \!LISTA untied}
         \psfrag{LAMP}[l][l][\sz]{\sf \!LAMP-l1 tied}
         \psfrag{LAMP untied}[l][l][\sz]{\sf \!LAMP-l1 untied}
         \psfrag{NMSE (dB)}[][][\szz]{\sf average NMSE [dB]}
         \psfrag{Layers}[][][\szz]{\sf layer / iteration}
         }

\putFrag{lista_vs_lamp_k15}
	{Test NMSE versus layer under $\vec{A}$ with condition number $15$.}
	{\figsize}
	{\newcommand{\sz}{0.53}
	 \newcommand{\szz}{0.7}
         \psfrag{LISTA}[l][l][\sz]{\sf \!LISTA tied}
         \psfrag{LISTA untied}[l][l][\sz]{\sf \!LISTA untied}
         \psfrag{LAMP}[l][l][\sz]{\sf \!LAMP-l1 tied}
         \psfrag{LAMP untied}[l][l][\sz]{\sf \!LAMP-l1 untied}
         \psfrag{NMSE (dB)}[][][\szz]{\sf average NMSE [dB]}
         \psfrag{Layers}[][][\szz]{\sf layer}
         }

\subsection{Effect of Shrinkage \texorpdfstring{$\vec{\eta}(\cdot)$}{eta} and Matrix \texorpdfstring{$\vec{A}$}{A}} 
\label{sec:numerical2}

Next we study the effect of the shrinkage choice $\vec{\eta}(\cdot)$ on network performance.
We begin by examining the performance of LAMP under the different shrinkage families proposed in \secref{shrinkage}.
In doing so, we will expose LAMP's lack of robustness to the matrix $\vec{A}$.
As a baseline, 
we also consider the \emph{support-oracle bound}, which is now described.
Suppose that an oracle provides knowledge of the support of $\vec{x}\true$. 
Then, since both the measurement noise $\vec{w}$ from \eqref{ys} and the
non-zero coefficients in $\vec{x}\true$ are Gaussian, 
the minimum MSE (MMSE) estimate of $\vec{x}\true$ from $\vec{y}$ can be computed in closed form.
This support-oracle MMSE lower bounds the MSE of any practical estimator of $\vec{x}\true$, which does not know the support.

\Figref{gaussianlamp} shows test-NMSE versus layer when the measurement matrix $\vec{A}$ is i.i.d.\ Gaussian. 
In the \emph{tied} case,
\figref{gaussianlamp} shows that the NMSEs achieved by LAMP with the BG, exponential, piecewise linear, and spline shrinkage functions are about $5$~dB better than those achieved by LAMP-$\ell_1$ (or, equivalently, LAMP with scaled-soft-threshold shrinkage).
Furthermore, the figure shows that there is relatively little difference in NMSE among the tied-LAMP networks with piecewise linear, exponential, spline, and BG shrinkage functions in this experiment.

\Figref{gaussianlamp} also shows that, for the BG and piecewise-linear shrinkages, the NMSE achieved by \emph{untied}\footnote{\Figref{gaussianlamp} shows untied LAMP performance with only piecewise linear and BG shrinkage functions, but the performance with exponential and spline shrinkage functions is very similar.} LAMP is about $1.5$~dB better than that of tied LAMP and only about $0.5$~dB away from the support-oracle bound after $10$ layers. 
The difference between untied LAMP and (tied) LISTA from \cite{Gregor:ICML:10} is remarkable, suggesting
that the combination of Onsager cancellation and optimized shrinkage is quite powerful.

\putFrag{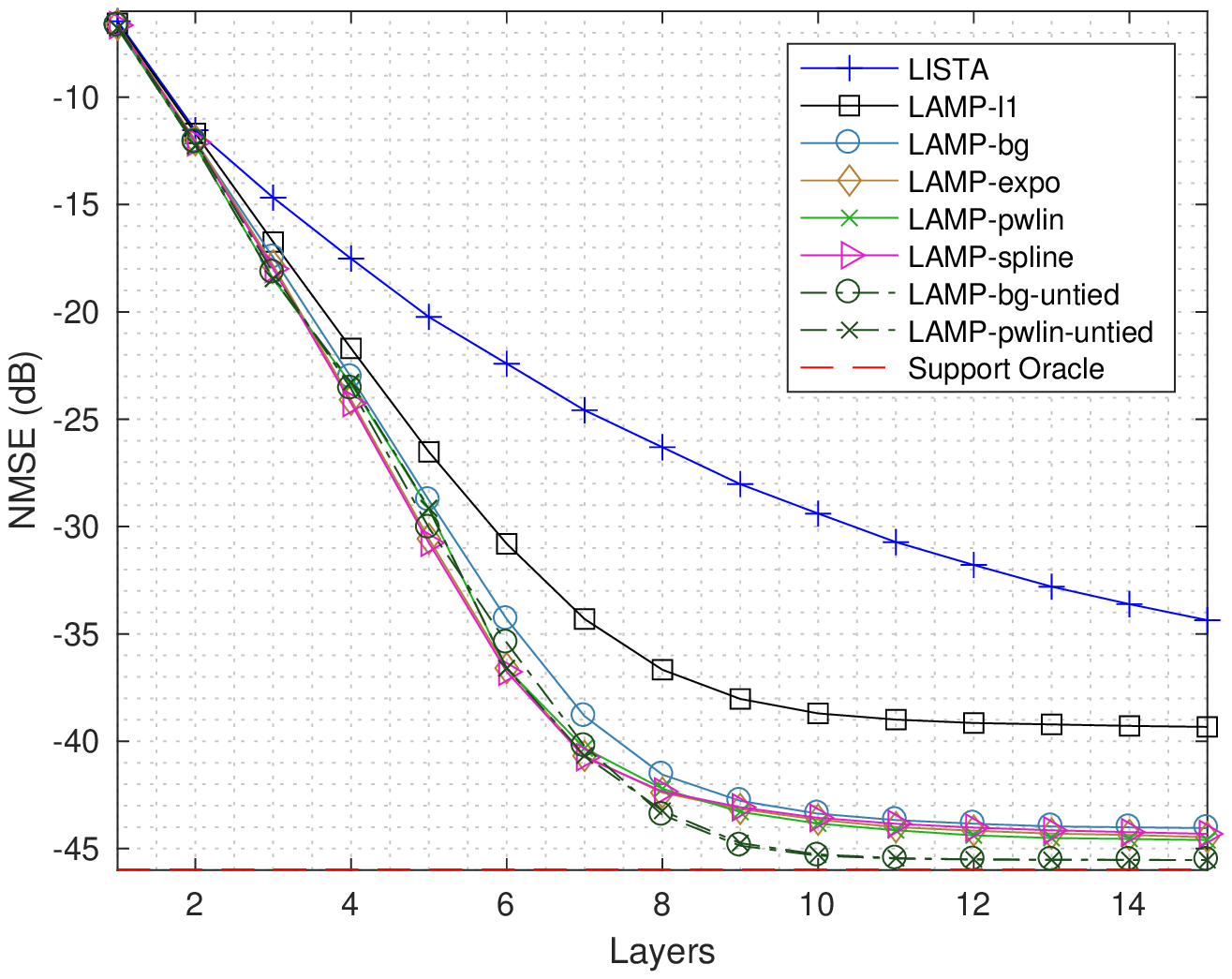}
	{Test NMSE versus layer under i.i.d.\ Gaussian $\vec{A}$.}
	{\figsize}
	{\newcommand{\sz}{0.59}
	 \newcommand{\szz}{0.7}
         \psfrag{LISTA}[l][l][\sz]{\sf \!LISTA tied}
         \psfrag{LAMP-l1}[l][l][\sz]{\sf \!LAMP-$\ell_1$ tied}
         \psfrag{LAMP-bg}[l][l][\sz]{\sf \!LAMP-bg tied}
         \psfrag{LAMP-pwlin}[l][l][\sz]{\sf \!LAMP-pwlin tied}
         \psfrag{LAMP-expo}[l][l][\sz]{\sf \!LAMP-exp tied}
         \psfrag{LAMP-spline}[l][l][\sz]{\sf \!LAMP-spline tied}
         \psfrag{LAMP-bg-untied}[l][l][\sz]{\sf \!LAMP-bg untied}
         \psfrag{LAMP-pwlin-untied}[l][l][\sz]{\sf \!LAMP-pwlin untied}
         \psfrag{Support Oracle}[l][l][\sz]{\sf \!support oracle}
         \psfrag{NMSE (dB)}[][][\szz]{\sf average NMSE [dB]}
         \psfrag{Layers}[][][\szz]{\sf layer}
         }

\Figref{kappa15lamp} shows test-NMSE versus layer when the measurement matrix $\vec{A}$ has condition number $\kappa=15$.
In the \emph{tied} case,
\figref{kappa15lamp} shows that the NMSEs achieved by LAMP with the BG, exponential, and spline shrinkage functions are about $5$~dB better than those achieved by LAMP-$\ell_1$, and that there is little difference among the NMSEs achieved by these shrinkage functions.
But, surprisingly, the piecewise linear shrinkage performs significantly better than the other shrinkages with $\geq 10$ layers and significantly worse with $< 10$ layers.

\putFrag{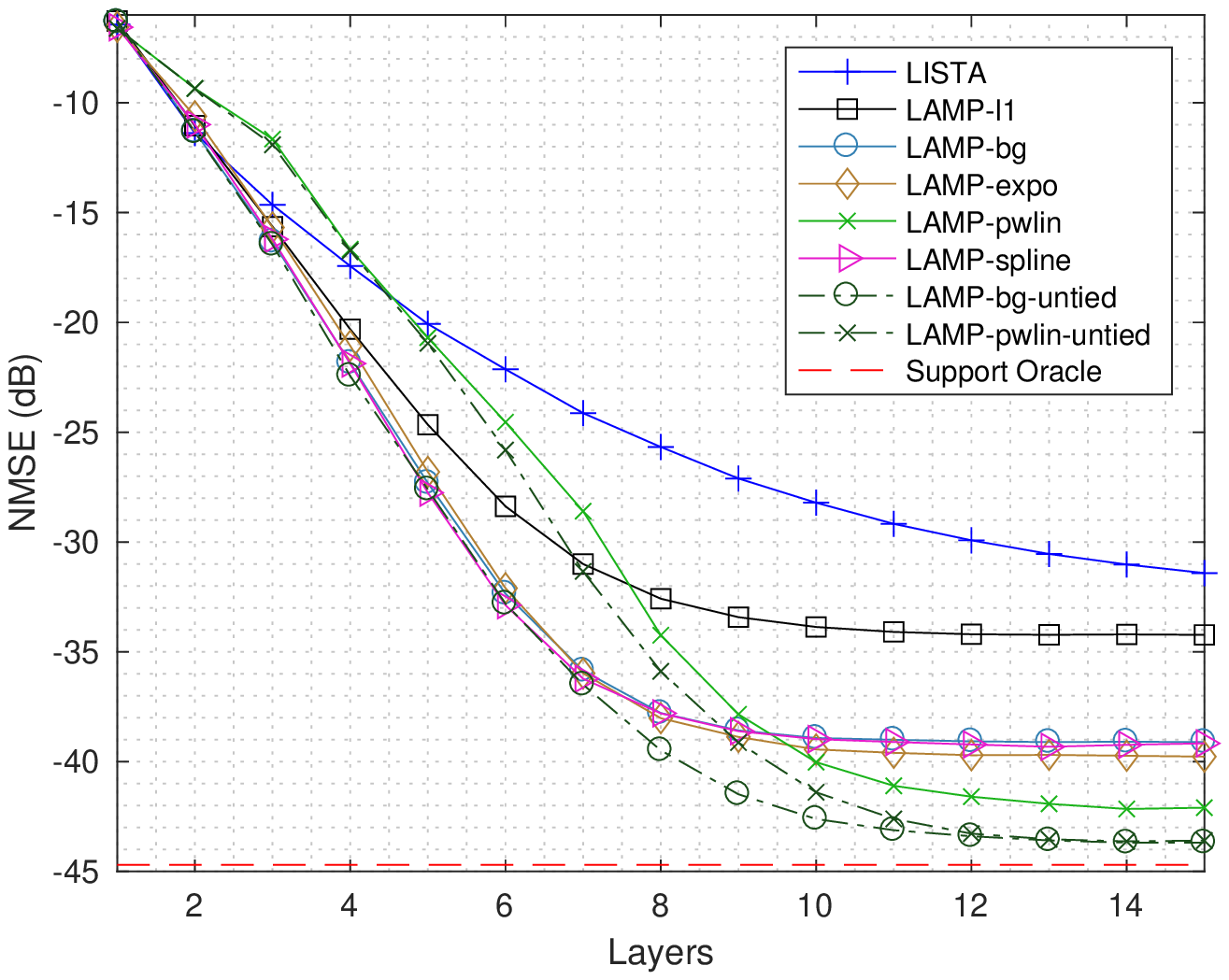}
	{Test NMSE versus layer under $\vec{A}$ with condition number $15$.}
	{\figsize}
        {\newcommand{\sz}{0.59}
	 \newcommand{\szz}{0.7}
         \psfrag{LISTA}[l][l][\sz]{\sf \!LISTA tied}
         \psfrag{LAMP-l1}[l][l][\sz]{\sf \!LAMP-$\ell_1$ tied}
         \psfrag{LAMP-bg}[l][l][\sz]{\sf \!LAMP-bg tied}
         \psfrag{LAMP-pwlin}[l][l][\sz]{\sf \!LAMP-pwlin tied}
         \psfrag{LAMP-expo}[l][l][\sz]{\sf \!LAMP-exp tied}
         \psfrag{LAMP-spline}[l][l][\sz]{\sf \!LAMP-spline tied}
         \psfrag{LAMP-bg-untied}[l][l][\sz]{\sf \!LAMP-bg untied}
         \psfrag{LAMP-pwlin-untied}[l][l][\sz]{\sf \!LAMP-pwlin untied}
         \psfrag{Support Oracle}[l][l][\sz]{\sf \!support oracle}
         \psfrag{NMSE (dB)}[][][\szz]{\sf average NMSE [dB]}
         \psfrag{Layers}[][][\szz]{\sf layer}
         }

With \emph{untied} LAMP, \figref{kappa15lamp} shows that BG shrinkage works very well: it dominates the other schemes at all layers $t$ and comes within $1$~dB of the support-oracle bound for $t\geq 13$~layers.
The piecewise-linear shrinkage works equally well with untied-LAMP for $t\geq 13$~layers, but significantly worse with fewer layers.

Together, Figs.~\ref{fig:gaussianlamp}-\ref{fig:kappa15lamp} suggest that LAMP behaves predictably with i.i.d.\ Gaussian $\vec{A}$, but less predictably with non-i.i.d.-Gaussian $\vec{A}$.
That is, since the true signal has a BG distribution, we would expect that the use of BG shrinkage would yield performance at least as good as other shrinkages and close to oracle bounds.
And this is precisely what happens with untied LAMP and i.i.d.\ Gaussian $\vec{A}$.
The fact that piecewise-linear shrinkage performs equally well under the same conditions can be explained by the fact that the piecewise-linear shrinkage function is flexible enough to mimic the BG shrinkage function.
But when $\vec{A}$ is not i.i.d.\ Gaussian, Figs.~\ref{fig:gaussianlamp}-\ref{fig:kappa15lamp} showed a strange gap in LAMP's performance with BG versus piecewise-linear shrinkages.
This suggests that LAMP might not be properly handling the non-i.i.d.\ Gaussian $\vec{A}$.
That said, LAMP is doing much better than AMP with this matrix, since AMP diverges.
We conjecture that the $\vec{B}$ matrix (or $\vec{B}_t$ matrices) learned by LAMP perform some sort of preconditioning that compensates for the non-i.i.d.-Gaussian singular-value spectrum of $\vec{A}$.

To further investigate the effect of measurement matrix $\vec{A}$, we examine the behavior of LAMP and (SVD-parameterized) LVAMP on a matrix $\vec{A}$ with condition number $\kappa=100$. 
(This matrix was constructed in the same way as the $\kappa=15$ matrix but with a different singular-value ratio $s_i/s_{i-1}=\rho$.)
\Figref{kappa100lvamp} shows that tied LAMP converges much more slowly with this $\kappa=100$ matrix; it takes many more layers for LAMP to attain a low NMSE.
Moreover, there is a huge gap between the BG and piecewise-linear versions of LAMP, which again suggests that LAMP is not properly handling the $\kappa=100$ matrix.
In contrast, \figref{kappa100lvamp} shows tied LVAMP converging in $15$ iterations to an NMSE that is not far from the oracle bound.
The proximity between LVAMP and matched VAMP in \figref{kappa100lvamp} is also interesting and will be discussed further below.

\putFrag{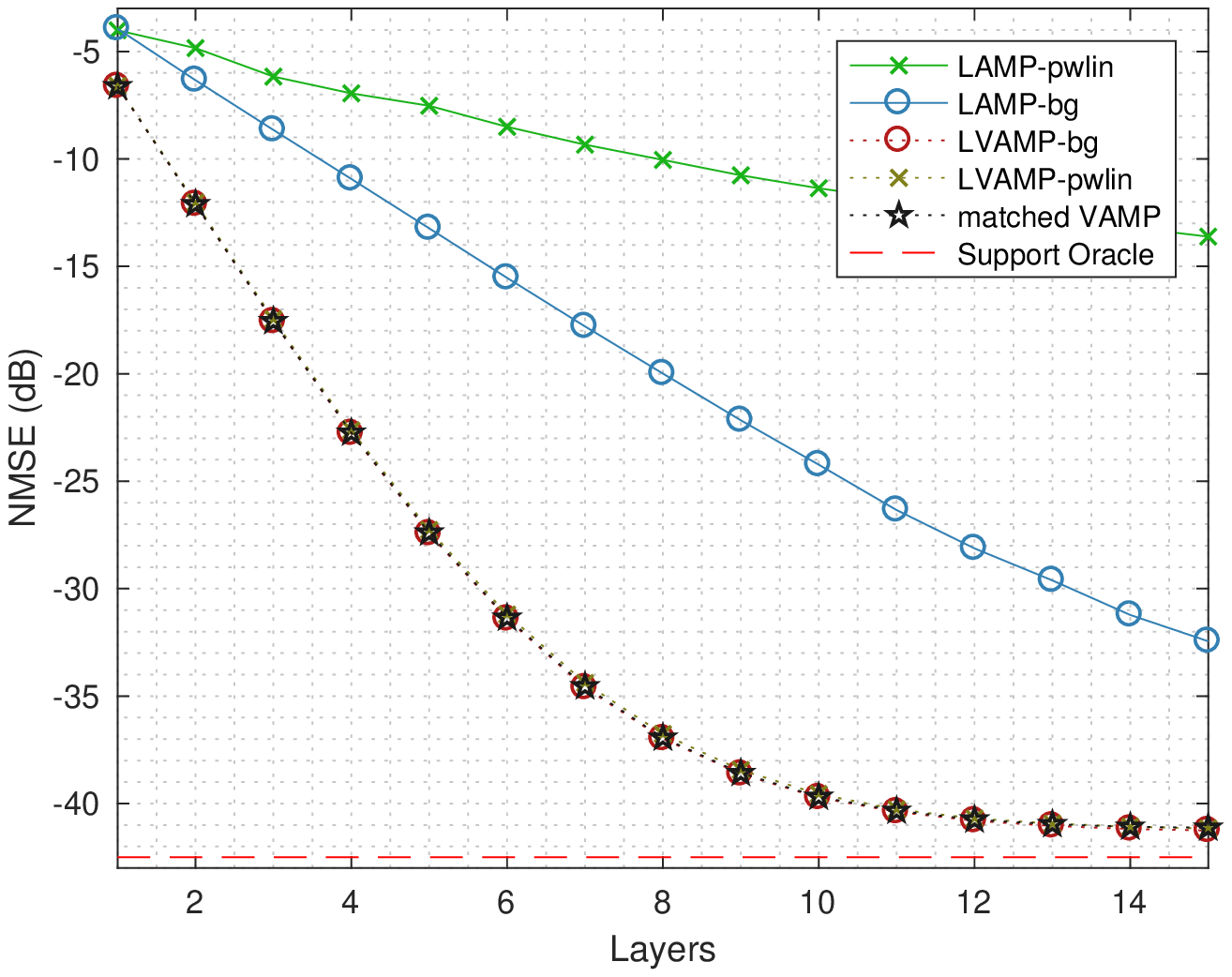}
    {Test NMSE versus layer (or versus iteration for matched VAMP) under $\vec{A}$ with condition number $100$.  The LVAMP traces represent both tied and untied SVD-parameter learning, which gave nearly identical results.}
    {\figsize}
    {\newcommand{\sz}{0.7}
     \newcommand{\szz}{0.56}
         \psfrag{LAMP-bg}[l][l][\szz]{\sf \!LAMP-bg tied}
         \psfrag{LAMP-pwlin}[Bl][Bl][\szz]{\sf \!LAMP-pwlin tied}
         \psfrag{LVAMP-bg}[l][l][\szz]{\sf \!LVAMP-bg}
         \psfrag{LVAMP-pwlin}[Bl][Bl][\szz]{\sf \!LVAMP-pwlin}
         \psfrag{Support Oracle}[l][l][\szz]{\sf \!support oracle}
         \psfrag{matched VAMP}[Bl][Bl][\szz]{\sf \!matched VAMP}
         \psfrag{NMSE (dB)}[][][\sz]{\sf average NMSE [dB]}
         \psfrag{Layers}[][][\sz]{\sf layer / iteration}
         }

\subsection{LVAMP's Robustness to the Matrix \texorpdfstring{$\vec{A}$}{A}} \label{sec:numerical3}

The experiments above suggest that LAMP performs well with i.i.d.-Gaussian $\vec{A}$, but that its convergence rate (in layers) slows as the matrix $\vec{A}$ becomes less well conditioned.
LVAMP, however, seems robust to ill-conditioning in $\vec{A}$, based on the results in \figref{kappa100lvamp}.
Thus, we now concentrate on evaluating LVAMP.
In doing so, we focus on the BG and piecewise-linear shrinkage families, for the reasons below. 
Because $\vec{x}\true$ is itself BG, the BG shrinkage should be optimal if the AWGN-corruption model \eqref{lsl2} holds.
But, in practice, we may not always know the distribution of the true signal, which motivates the use of a flexible shrinkage family, like exponential, spline, or piecewise linear. 
Among those, our previous experiments showed that piecewise-linear shrinkage exposed weaknesses in the LAMP framework, although it performed well after many layers. 
Thus, we focus on the BG and piecewise-linear shrinkages when evaluating LVAMP.

Figures~\ref{fig:gaussianlvamp}-\ref{fig:kappa15lvamp} show test-NMSE versus layer for i.i.d.\ Gaussian $\vec{A}$ and $\vec{A}$ with condition number $\kappa=15$, respectively.
In those figures, ``LVAMP'' refers to both the \emph{tied} and \emph{untied} versions of LVAMP, which gave essentially identical NMSE.
In fact, the connection is even stronger:
at every layer $t$, the values of the parameters $(\tvec{\theta}_t,\vec{\theta}_t)$ learned by untied LVAMP were nearly identical to the values of the parameters $(\tvec{\theta},\vec{\theta}_t)$ learned by tied LVAMP (where here $\tvec{\theta}=\{\vec{U},\vec{s},\vec{V},\sigma_w^2\}$).
We will discuss this connection further in the \secref{numerical4}.

\Figref{gaussianlvamp} shows test-NMSE versus layer when the measurement matrix $\vec{A}$ is i.i.d.\ Gaussian. 
There, we first notice that NMSE of LVAMP is about $2$~dB better than that of tied LAMP for networks with $>4$ layers, for both BG and piecewise-linear shrinkage.
Second, the NMSE of LVAMP is noticeably better than of \emph{untied} LAMP for networks with $4$-$8$ layers.
But, with $>10$ layers, the two schemes perform equally well and within $0.5$~dB of the support-oracle bound.

\Figref{kappa15lvamp} shows test-NMSE versus layer when the measurement matrix $\vec{A}$ has condition number $\kappa(\vec{A})=15$.
There, we first notice that NMSE of LVAMP is $2$-$5$~dB better than that of tied LAMP for networks with $>4$ layers, for both BG and piecewise-linear shrinkage.
Second, the NMSE of LVAMP is $0.5$-$2$~dB better than of \emph{untied} LAMP at all layers and within $0.5$~dB of the support-oracle bound for $\geq 10$ layers.

\putFrag{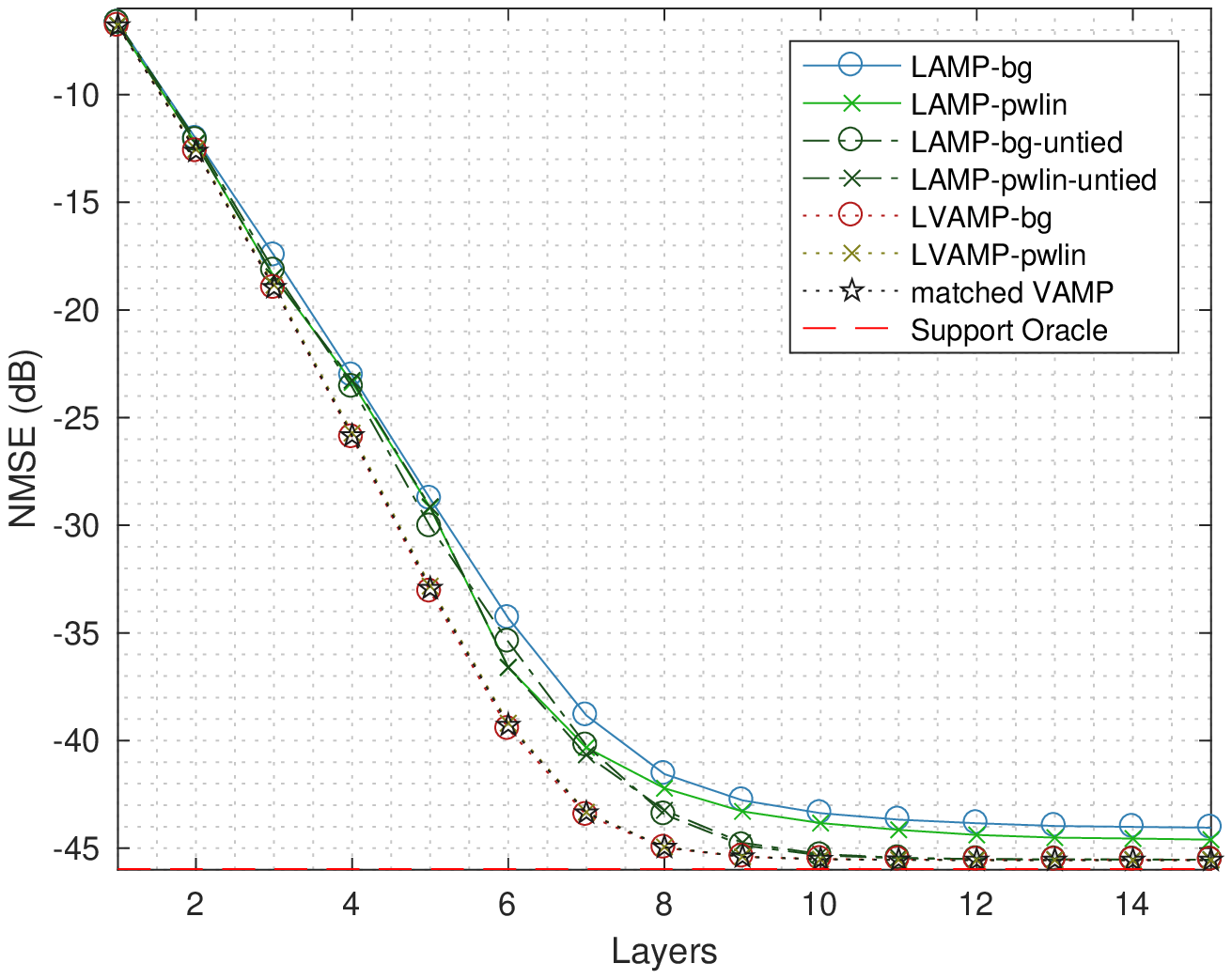}
    {Test NMSE versus layer (or versus iteration for matched VAMP) under i.i.d.\ Gaussian $\vec{A}$.  The LVAMP traces represent both tied and untied SVD-parameter learning, which gave nearly identical results.}
    {\figsize}
    {\newcommand{\sz}{0.7}
     \newcommand{\szz}{0.6}
         \psfrag{LAMP-bg}[l][l][\szz]{\sf \!LAMP-bg tied}
         \psfrag{LAMP-pwlin}[Bl][Bl][\szz]{\sf \!LAMP-pwlin tied}
         \psfrag{LAMP-bg-untied}[l][l][\szz]{\sf \!LAMP-bg untied}
         \psfrag{LAMP-pwlin-untied}[Bl][Bl][\szz]{\sf \!LAMP-pwlin untied}
         \psfrag{LVAMP-bg}[l][l][\szz]{\sf \!LVAMP-bg}
         \psfrag{LVAMP-pwlin}[Bl][Bl][\szz]{\sf \!LVAMP-pwlin}
         \psfrag{matched VAMP}[Bl][Bl][\szz]{\sf \!matched VAMP}
         \psfrag{Support Oracle}[l][l][\szz]{\sf \!support oracle}
         \psfrag{NMSE (dB)}[][][\sz]{\sf average NMSE [dB]}
         \psfrag{Layers}[][][\sz]{\sf layer / iteration}
         }

\putFrag{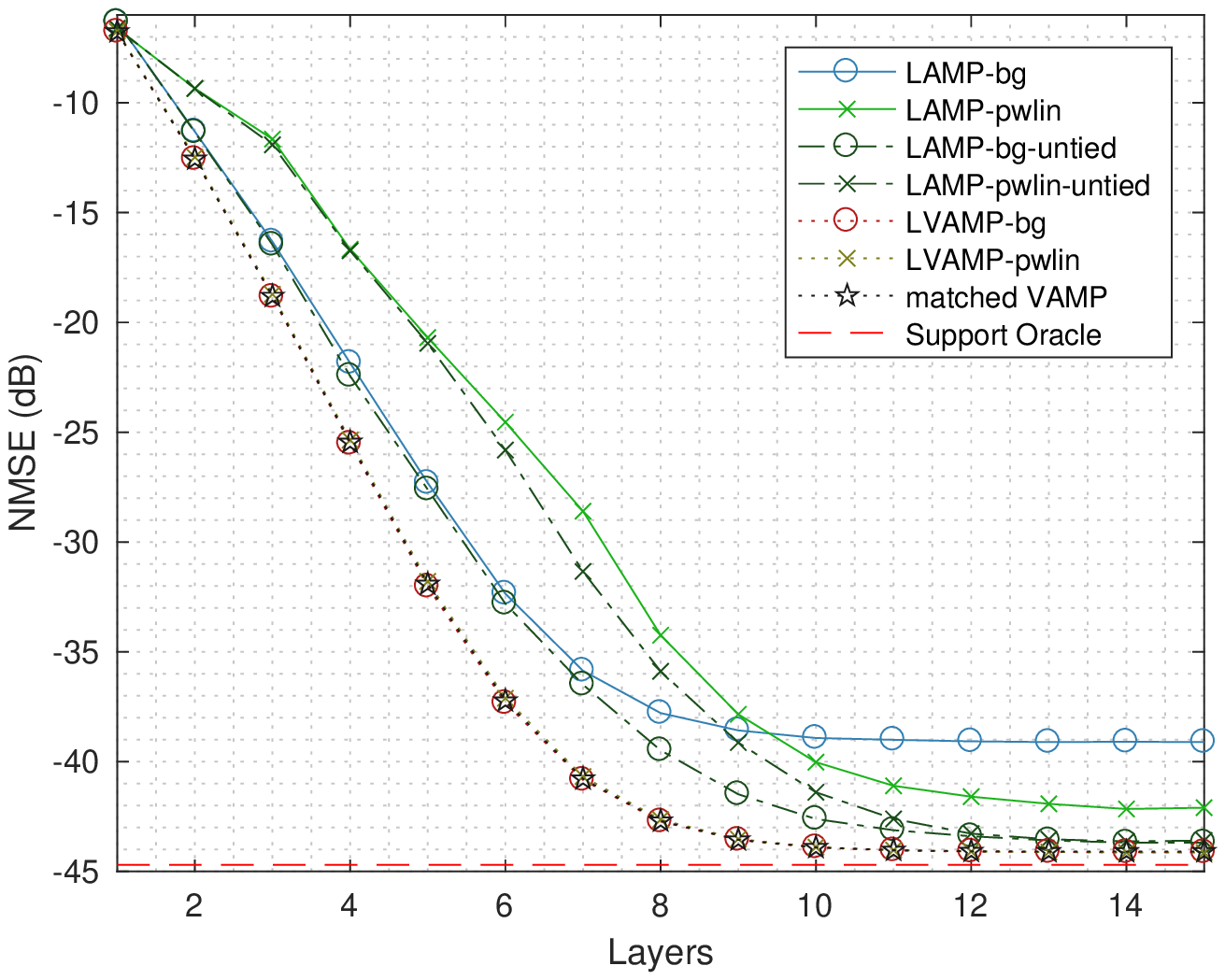}
    {Test NMSE versus layer (or versus iteration for matched VAMP) under $\vec{A}$ with condition number $15$.  The LVAMP traces represent both tied and untied SVD-parameter learning, which gave nearly identical results.}
    {\figsize}
    {\newcommand{\sz}{0.7}
     \newcommand{\szz}{0.6}
         \psfrag{LAMP-bg}[l][l][\szz]{\sf \!LAMP-bg tied}
         \psfrag{LAMP-pwlin}[Bl][Bl][\szz]{\sf \!LAMP-pwlin tied}
         \psfrag{LAMP-bg-untied}[l][l][\szz]{\sf \!LAMP-bg untied}
         \psfrag{LAMP-pwlin-untied}[Bl][Bl][\szz]{\sf \!LAMP-pwlin untied}
         \psfrag{LVAMP-bg}[l][l][\szz]{\sf \!LVAMP-bg}
         \psfrag{LVAMP-pwlin}[Bl][Bl][\szz]{\sf \!LVAMP-pwlin}
         \psfrag{matched VAMP}[Bl][Bl][\szz]{\sf \!matched VAMP}
         \psfrag{Support Oracle}[l][l][\szz]{\sf \!support oracle}
         \psfrag{NMSE (dB)}[][][\sz]{\sf average NMSE [dB]}
         \psfrag{Layers}[][][\sz]{\sf layer / iteration}
         }

Looking at Figs.~\ref{fig:kappa100lvamp}-\ref{fig:kappa15lvamp} together, we see that the advantage of LVAMP over untied-LAMP is relatively small for i.i.d.\ Gaussian $\vec{A}$ but grows with the condition number of $\vec{A}$.
We also see that, with LVAMP, there is essentially no difference in the performance of BG shrinkage versus piecewise-linear shrinkage for any $\vec{A}$.

\subsection{Equivalence of LVAMP and Matched VAMP} \label{sec:numerical4}

Perhaps the most interesting behavior in Figures~\ref{fig:kappa100lvamp}-\ref{fig:kappa15lvamp} is the following. 
\emph{The NMSEs achieved by the LVAMP networks are indistinguishable from those of the matched VAMP algorithm} (i.e., VAMP under statistically matched i.i.d.\ signal and noise models) for all $\vec{A}$ under test.
And looking at the parameters $\{\tvec{\theta}_t,\vec{\theta}_t\}$, where $\tvec{\theta}_t=\{\vec{U}_t,\vec{s}_t,\vec{V}_t,\sigma_{wt}^2\}$, those \emph{learned} by LVAMP-BG%
\footnote{For the LVAMP traces in Figures~\ref{fig:kappa100lvamp}-\ref{fig:kappa15lvamp}, we did not use an $\{\vec{U},\vec{s},\vec{V}\}$ initialization that matched the SVD of $\vec{A}$, as recommended in \secref{lvamplearn}. Rather, the $\{\vec{U},\vec{s},\vec{V}\}$ initialization was chosen randomly, to test if back-propagation would learn the matched values.}
coincide almost perfectly with those \emph{prescribed} by matched VAMP.
In this sense, matched VAMP ``predicts'' the parameters learned by back-propagation.

But, beyond merely a prediction, matched VAMP offers an \emph{explanation} of the parameters learned by LVAMP.
Recall that the $t$th iteration of matched VAMP comprises four operations:
1) MSE-optimal vector estimation of $\vec{x}$ 
from measurements $\vec{y}=\vec{Ax}+\mc{N}(\vec{0},\sigma_w^2\vec{I})$
and pseudo-prior $\vec{x}\sim\mc{N}(\tvec{r}_t,\tilde{\sigma}_t^2\vec{I})$,
2) an Onsager decoupling stage that yields 
the pseudo-measurement $\vec{r}_t=\vec{x}+\mc{N}(\vec{0},\sigma_t^2\vec{I})$,
3) MSE-optimal scalar estimation of i.i.d.\ $\vec{x}$ 
under pseudo-measurement $\vec{r}_t$ and prior $\vec{x}\sim \prod_j p_j(x_j)$,
and
4) an Onsager decoupling stage that yields 
the pseudo-prior parameters $(\tvec{r}_t,\tilde{\sigma}_t^2)$.
From this understanding of matched VAMP, it follows that 
the linear stage of LVAMP learns parameters $\tvec{\theta}_t$ that are MSE-optimal under the pseudo-prior $\vec{x}\sim\mc{N}(\tvec{r}_t,\tilde{\sigma}_t^2\vec{I})$ generated by the preceding Onsager decoupling stage.
Likewise, the nonlinear stage of LVAMP learns shrinkage-function parameters $\vec{\theta}_t$ that are MSE-optimal under the pseudo-measurements $\vec{r}_t=\vec{x}+\mc{N}(\vec{0},\sigma_t^2\vec{I})$ generated by the preceding Onsager decoupling stage.

From a practical standpoint, the significance of the agreement between LVAMP-BG and matched VAMP is somewhat diminished by the fact that both approaches used knowledge of the prior family on $\vec{x}\true$ (in this case, BG). 
But LVAMP with piece-linear shrinkage performed just as well as matched VAMP in Figures~\ref{fig:kappa100lvamp}-\ref{fig:kappa15lvamp}.  
And, for piecewise-linear shrinkage, no knowledge of the prior on $\vec{x}\true$ was used (beyond i.i.d.).

\section{Application to 5G Communications} \label{sec:5G}

In this section we demonstrate the application of our proposed methods to two important problems from 5th-generation (5G) wireless communications \cite{Andrews:JSAC:14,Wunder:ACC:15}: 
\emph{compressive random access} and \emph{massive-MIMO channel estimation}.
As we describe in the sequel, both can be posed as instances of the sparse linear inverse problem described in \secref{intro}.
For LVAMP, we used the LMMSE parameterization \eqref{lmmse3}.

\subsection{Application to Compressive Random Access} \label{sec:CRA}

5G communications systems will need to support the ``internet of things,'' which will bring billions of everyday objects (e.g., light bulbs, washer/dryers, ovens, etc.) online. 
Since these devices will connect only sporadically and often have little data to communicate when they do, it is important that they can access the system with little control overhead.

Towards this aim, it has been suggested to assign, to each user (i.e., device) in a given cell, a unique length-$M$ pilot sequence.
When a user wants to connect the base station (BS), it waits to hear a synchronization beacon emitted by the BS and then broadcasts its pilots.
The signal $\vec{y}$ received by the BS then takes the form in \eqref{y}, where 
the $n$th column of $\vec{A}$ is the pilot sequence of the $n$th user; 
the $n$th entry of $\vec{x}\true$ is determined by 
the activity of the $n$th user (i.e., $x_n\true=0$ if inactive)
as well as its propagation channel to the BS; 
and the vector $\vec{w}$ models out-of-cell interference and thermal noise.
(See the detailed model in \appref{5G}.)
Assuming that users are sporadically connected, the $\vec{x}\true$ vector will be sparse, allowing the use of sparse signal recovery for 
\emph{joint user-activity detection and channel estimation}
\cite{Fletcher:ISIT:09,Bockelmann:TETT:13,Wunder:ACC:15}.

If the pilots $\vec{A}$ are drawn i.i.d.\ Gaussian
and the number of users $N$ is large,
then the support-recovery analysis from \cite[Corollary 2]{Wainwright:TIT:09} says that, 
in order to accurately\footnote{By ``accurately'' we mean that the probability of detection error converges to zero as $N\rightarrow\infty$ \cite{Wainwright:TIT:09}.}
detect the active subset of $N$ users 
under activity rate $\gamma\in (0,1)$,
the $\ell_1$ approach \eqref{lasso} 
requires pilots of length $M > 2\gamma N \log[(1-\gamma) N]$.
For example, with $N=512$ users and activity rate $\gamma=0.01$, this analysis suggests to use pilots of length $M \geq 64$.
Because $M\ll N$ and the user activities are random, this formulation is often referred to as ``compressive random access.''

We now numerically investigate the performance of LAMP and LVAMP on the compressive random access problem described above.
For our experiment, we assumed that the pilots in $\vec{A}$ were i.i.d.\ QPSK (i.e., uniformly distributed over $\{\j,1,-\j,-1\}$, where $\j\defn\sqrt{-1}$).
Such pilots are common, as they result in low peak-to-average power ratio at the transmitter.
Also, we assumed that the activity/channel coefficients $\vec{x}\true$ were distributed as described in \appref{5G}, assuming users uniformly distributed over a single hexagonal cell with a one-antenna BS (for simplicity). 
Finally, we assumed AWGN $\vec{w}$ with power adjusted to achieve SNR $=10$~dB.
For training, we used a single realization of $\vec{A}\in\Complex^{M\times N}$ and $1024$ random draws of $\vec{x}\true\in\Complex^N$ for each mini-batch,
and for testing we used the same $\vec{A}$ and $1024$ new random draws of $\vec{x}\true$.
Finally, we assumed $N=512$ users, activity rate $\gamma=0.01$, and---inspired by the $\ell_1$-analysis from \cite{Wainwright:TIT:09}---pilots of length $M=64$.

\Figref{singleantennasinglecell} shows test-NMSE versus layer for the compressive random access problem described above.  
There we see that the LAMP and LVAMP methods significantly outperformed both tied and untied LISTA.
For both the LAMP and LVAMP methods, the piecewise linear shrinkage performed about $0.5$~dB better than the BG shrinkage,
untied LVAMP performed about $0.5$~dB better than untied LAMP,
and untied LAMP performed about $0.5$~dB better than tied LAMP.
We conjecture that the small difference between untied LAMP and LVAMP is due to the i.i.d.\ property of the matrix $\vec{A}$.

\putFrag{singleantennasinglecell}
    {Test NMSE versus layer for compressive random access.  LVAMP used the LMMSE parameterization \eqref{lmmse3}.}
    {\figsize}
    {\newcommand{\sz}{0.7}
     \newcommand{\szz}{0.6}
         \psfrag{LISTA}[l][l][\szz]{\sf \!LISTA tied}
         \psfrag{LISTAuntied}[l][l][\szz]{\sf \!LISTA untied}
         \psfrag{LAMPbg}[l][l][\szz]{\sf \!LAMP-bg tied}
         \psfrag{LVAMPbg}[l][l][\szz]{\sf \!LVAMP-bg untied}
         \psfrag{LAMPpwgrid}[Bl][Bl][\szz]{\sf \!LAMP-pwlin tied}
         \psfrag{LAMPpwgriduntied}[Bl][Bl][\szz]{\sf \!LAMP-pwlin untied}
         \psfrag{LVAMP-pwgrid}[Bl][Bl][\szz]{\sf \!LVAMP-pwlin untied}
         \psfrag{LAMP-bg-untied}[l][l][\szz]{\sf \!LAMP-bg}
         \psfrag{matched VAMP}[Bl][Bl][\szz]{\sf \!matched VAMP}
         \psfrag{Support Oracle}[l][l][\szz]{\sf \!support oracle}
         \psfrag{NMSE (dB)}[][][\sz]{\sf average NMSE [dB]}
         \psfrag{Layers}[][][\sz]{\sf layer}
         }

\subsection{Application to Massive-MIMO Channel Estimation} \label{sec:MIMO}

So-called ``massive-MIMO'' \cite{Rusek:SPM:13} is likely to play a large role in 5G wireless \cite{Andrews:JSAC:14}. 
In such systems, the BS has a massive antenna array (i.e., dozens or hundreds of elements) and the user devices have single antennas.
The idea is that, by making the number of BS antennas $\Nr$ very large, the array gain becomes very large, which then drives both (in-cell) multiuser interference and thermal noise to very low levels.
But doing so requires accurate channel-state information (CSI).

To obtain this CSI, it is envisioned that the users will simultaneously broadcast known pilots, which the BS will use to estimate the uplink channels.
Through time-division duplex and channel reciprocity, the same estimates can be used for the downlink.
The main bottleneck in such systems results from ``pilot contamination'' \cite{Rusek:SPM:13}.
That is, the pilots used in a given cell may be the same as those used in a neighboring cell, which results in contaminated channel estimates and hence out-of-cell interference that does not vanish as $\Nr$ increases.

One way to circumvent pilot contamination is to assign random pilots in every cell and estimate both the in- and out-of-cell user channels at each BS (assuming knowledge of the pilots in neighboring cells) \cite{Wen:TWC:15}.
Although the computational complexity of such an approach 
may seem high, it can be reduced by processing each (of the $\Nr$) receive angles separately. 
Because relatively few users contribute significant energy to a given receive angle, the per-angle channel coefficients are approximately sparse.
The resulting channel-estimation problem takes the form of of \eqref{y}, where now $\vec{y}$ represents the temporal measurements for a given receive angle, $\vec{A}\in\Complex^{M\times N}$ the pilots, $\vec{x}\true$ the per-angle channel coefficients, and $\vec{w}$ thermal noise.
Finally, $M$ represents the pilot duration and $N$ represents the total number of users in the primary and neighboring cells.
(See \appref{5G} for details.)

We now numerically investigate the performance of LAMP and LVAMP on the massive-MIMO channel-estimation problem described above.
For this, we assumed i.i.d.\ QPSK pilots $\vec{A}$; 
$1$ primary cell and $6$ interfering cells (all hexagonal)
with $64$ users uniformly distributed within each cell (so that $N=7\times 64=448$); 
pilot sequences of length $M=64$; 
$\Nr=64$ BS antennas;
and an SNR of $20$~dB.
Channels $\vec{x}\true$ were generated as described in \appref{5G} and $\vec{w}$ was AWGN.
Different from our random-access formulation,
all users transmit pilots,
and $\vec{w}$ does not model interference from nearby cells
(yielding higher SNR$\defn\E\{\|\vec{Ax}\true\|^2\}/\E\{\|\vec{w}\|^2\}$).

\Figref{multiantennamulticell} shows test-NMSE versus layer for the massive-MIMO channel estimation problem described above, where NMSE is measured only on the channels of primary-cell users.
The results in the figure look as expected: for piecewise-linear shrinkage, the ranking (from best to worst at 6 layers) is LVAMP, untied LAMP, tied LAMP, untied LISTA, and tied LISTA.
Meanwhile, piecewise-linear shrinkage outperformed BG shrinkage by roughly $0.5$~dB.
We conjecture that the small difference between untied LAMP and LVAMP is due to the i.i.d.\ property of the matrix $\vec{A}$.

\putFrag{multiantennamulticell}
    {Test NMSE versus layer for massive-MIMO channel estimation.  LVAMP used the LMMSE parameterization \eqref{lmmse3}.}
    {\figsize}
    {\newcommand{\sz}{0.7}
     \newcommand{\szz}{0.6}
         \psfrag{LISTA}[l][l][\szz]{\sf \!LISTA tied}
         \psfrag{LISTAuntied}[l][l][\szz]{\sf \!LISTA untied}
         \psfrag{LAMPbg}[l][l][\szz]{\sf \!LAMP-bg tied}
         \psfrag{LVAMPbg}[l][l][\szz]{\sf \!LVAMP-bg untied}
         \psfrag{LAMPpwgrid}[Bl][Bl][\szz]{\sf \!LAMP-pwlin tied}
         \psfrag{LAMPpwgriduntied}[Bl][Bl][\szz]{\sf \!LAMP-pwlin untied}
         \psfrag{LVAMP-pwgrid}[Bl][Bl][\szz]{\sf \!LVAMP-pwlin untied}
         \psfrag{LAMP-bg-untied}[l][l][\szz]{\sf \!LAMP-bg}
         \psfrag{matched VAMP}[Bl][Bl][\szz]{\sf \!matched VAMP}
         \psfrag{Support Oracle}[l][l][\szz]{\sf \!support oracle}
         \psfrag{NMSE (dB)}[][][\sz]{\sf average NMSE [dB]}
         \psfrag{Layers}[][][\sz]{\sf layer}
         }

\subsection{Discussion}

We also tried implementing a convolutional neural network (CNN) to solve the two 5G problems above, but we did not obtain good results.
In particular, we tried an implementation of the DeepInverse approach from \cite{Mousavi:ICASSP:17}.
Although CNNs give state-of-the-art performance in image recovery, they do not appear to be well suited to problems where there is little structure other than sparsity.
Conversely, CNNs are known to work very effectively in recovering richly structured signals, such as images, where our preliminary experiments with LAMP and LVAMP have not show state-of-the-art results.

\section{Conclusion} 

In this paper, we proposed two deep-learning approaches to the sparse linear inverse problem described in \secref{intro}.
Our first approach, LAMP, is obtained by unfolding the AMP algorithm \cite{Donoho:PNAS:09} into a deep network and learning the network parameters that best fit a large training dataset.
Although reminiscent of Gregor and LeCun's LISTA \cite{Gregor:ICML:10}, it differs in i) the inclusion of Onsager correction paths that decouple errors across layers, and ii) joint learning of the linear transforms and nonlinear shrinkage functions.
To avoid convergence to bad local minima, we proposed a reparameterization of AMP and a hybrid layer-wise/global learning strategy.
Our second approach, LVAMP, is obtained by unfolding the VAMP algorithm \cite{Rangan:VAMP} into a deep network and learning its linear and nonlinear parameters using similar methods.

A synthetic numerical study showed that LAMP and LVAMP significantly outperformed LISTA in both convergence rate (in layers) and final MSE.
And, while the performance of LAMP deteriorated with ill-conditioning in the matrix $\vec{A}$, that for LVAMP did not.
Interestingly, with i.i.d.\ signals, the network parameters learned by LVAMP were nearly identical to the ones prescribed by the matched VAMP algorithm, i.e., VAMP with statistically matched prior and likelihood.
Thus, the MMSE-estimation principles that underlie VAMP offer an intuitive interpretation of LVAMP. 

We also applied LAMP and LVAMP to two problems in 5G wireless communications: compressive random access and massive-MIMO channel estimation, 
where we saw gains relative to LISTA and more conventional deep CNNs.
We conjecture that, for image recovery applications, it would be more appropriate to unfold and learn a \emph{multi-layer} AMP \cite{Manoel:17} or VAMP algorithm, which is a topic of ongoing work.

We also see value in extending the LAMP and LVAMP methods from the linear model \eqref{y} to the \emph{generalized linear} model $\vec{y}=f(\vec{Ax}+\vec{w})$, where $f(\cdot)$ is a known, componentwise nonlinearity.
For this, it may be possible to unfold the \emph{generalized} AMP \cite{Rangan:ISIT:11} and VAMP \cite{Schniter:ASIL:16} algorithms into networks and learn improved network parameters from training data.
Doing so would facilitate the application of AMP-inspired deep networks to problems such as 
phase retrieval \cite{Schniter:TSP:15} and 
quantized compressive sensing \cite{Kamilov:TSP:12}.

\appendices
\section{Derivation of LAMP-\texorpdfstring{$\ell_1$}{L1} Equations \eqref{lamp}} \label{app:redefine}

From \figref{lamp_1layer}, the $t$th layer of LAMP implements
\begin{subequations}
\label{eq:lampB}
\begin{align}
\hvec{x}_{t+1}
&= \shrinkage{st}\left(\hvec{x}_t+\vec{B}_t\vec{v}_t;\lambda_t\right) \\
\vec{v}_{t+1}
&= \vec{y} - \vec{A}_t\hvec{x}_{t+1} + b_{t+1} \vec{v}_t .
\end{align}
\end{subequations}
Substituting \eqref{simplify} into \eqref{lampB} gives 
\begin{subequations}
\label{eq:lampC}
\begin{align}
\hvec{x}_{t+1}
&= \shrinkage{st}\left(\hvec{x}_t+\vec{B}_t\vec{v}_t;\lambda_t\right) \\
\vec{v}_{t+1}
&= \vec{y} - \beta_t\vec{A}\hvec{x}_{t+1} + b_{t+1} \vec{v}_t .
\end{align}
\end{subequations}
Defining $\ovec{x}_t\defn\beta_t\hvec{x}_t$ and $\ovec{B}_t\defn\beta_t\vec{B}_t$, we can write \eqref{lampC} as
\begin{subequations}
\label{eq:lampD}
\begin{align}
\ovec{x}_{t+1}
&= \beta_{t+1}\shrinkage{st}\left(\frac{\ovec{x}_t+\ovec{B}_t\vec{v}_t}{\beta_t};\lambda_t\right) 
\label{eq:xbar}\\
\vec{v}_{t+1}
&= \vec{y} - \vec{A}\ovec{x}_{t+1} + b_{t+1} \vec{v}_t 
\label{eq:v} .
\end{align}
\end{subequations}
Since the soft thresholder \eqref{soft_thresh} obeys
$\shrinkage{st}(\vec{r};\lambda)=\shrinkage{st}(\beta\vec{r};\beta\lambda)/\beta$
for any $\beta>0$,
equation \eqref{xbar} can be written as
\begin{align}
\ovec{x}_{t+1}
&= \frac{\beta_{t+1}}{\beta_t}\shrinkage{st}\left(\ovec{x}_t+\ovec{B}_t\vec{v}_t;\beta_t\lambda_t\right) \\
&= \bar{\beta}_t\shrinkage{st}\left(\ovec{x}_t+\ovec{B}_t\vec{v}_t;\bar{\lambda}_t\right) 
\label{eq:xbar2} ,
\end{align}
where $\bar{\beta}_t\defn \beta_{t+1}/\beta_t$ and 
$\bar{\lambda}_t\defn\beta_t\lambda_t$.
Finally, using the definitions of $\lambda_t$ and $b_t$ from \eqref{lambda} and \eqref{btgen}, and defining $\bar{\alpha}_t\defn \beta_t\alpha_t$,
equations \eqref{v} and \eqref{xbar2} imply that the $t$th layer of LAMP implements
\begin{subequations}
\label{eq:lampE}
\begin{align}
\ovec{x}_{t+1}
&= \bar{\beta}_t\shrinkage{st}\left(\ovec{x}_t+\ovec{B}_t\vec{v}_t;\tfrac{\bar{\alpha}_t}{\sqrt{M}}\|\vec{v}_t\|_2\right) \\
\vec{v}_{t+1}
&= \vec{y} - \vec{A}\ovec{x}_{t+1} + \tfrac{\bar{\beta}_t}{M}\|\ovec{x}_{t+1}\|_0 \vec{v}_t ,
\end{align}
\end{subequations}
where $\ovec{B}_t, \bar{\beta}_t, \bar{\alpha}_t$ are freely adjustable parameters.
To avoid an overabundance of notation in the main body of the paper, we rewrite \eqref{lampE} as \eqref{lamp} by redefining $\hvec{x}_t\leftarrow\ovec{x}_t$ and dropping the bars on the remainder of the variables.

\section{5G Channel Modeling Details} \label{app:5G}

In this section we provide details for the system model used in \secref{5G}.
To save space, we present a general model that yields both compressive random access and massive-MIMO channel estimation as special cases.

Consider a wireless system with $\Nc$ nearby cells, where each cell contains up to $\Nu$ single-antenna users and a BS with $\Nr$ antennas.
Each BS is assumed to use a uniform linear array with half-wavelength element spacing.
The BSs are time-synchronized and periodically broadcast a beacon.
Upon hearing the beacon, the active users simultaneously broadcast pilot waveforms that reach each BS through multipath propagation. 
The BS of interest will then measure, at discrete time $m=1\dots M$ and antenna $q=1\dots\Nr$,
\begin{eqnarray}
[\ovec{Y}]_{mq} 
&=& \sum_{n=1}^{N} \sum_{p=1}^{P_n} 
        \delta_n a_n(mT-\tau_{np}) g_{np} e^{\j \theta_{np}q} 
        + [\ovec{W}]_{mq} ,
\quad \label{eq:Ybar}
\end{eqnarray}
where, for user $n=1\dots\Nc\Nu$, the quantity
$\delta_n\in\{0,1\}$ is the activity indicator;
$a_n(t)$ is the pilot waveform;
$P_n$ are the number of propagation paths;
$g_{np}$, $\tau_{np}$, and $\theta_{np}$ are the gain, delay, and arrival angle of the $p$th path;
$T$ is the sampling interval;
and 
$\ovec{W}$ is noise and residual interference from far-away cells.

We model the path gain/loss as $g_{np}=h_{np}/(1+d_n^\rho)$, where $d_n$ is the distance from the $n$th user to the BS, $\rho$ is the path-loss exponent, and $h_{np}$ is a random fluctuation such that $\E\{\sum_p |h_{np}|^2\}=1$. 
In our experiments, we used 
$P_n=5$ paths with angle spread $10^\circ$ 
and Rician fading with k-factor $10$ for $h_{np}$, 
and we used $\rho=4$ for the path-loss exponent.

We assume that the waveforms $a_n(t)$ are approximately bandlimited to $T^{-1}$~Hz and---for simplicity---that $\tau_{np}\ll T$, yielding the ``narrowband'' approximation
$a_n(mT-\tau_{np})\approx a_n(mT)\defn a_{mn}$, so that
\begin{align}
[\ovec{Y}]_{mq} 
&= \sum_{n=1}^N a_{mn} z_{nq} + [\ovec{W}]_{mq} ,
\label{eq:Ybar2}
\end{align}
for
$N\defn \Nc\Nu$
and
$z_{nq} \defn \delta_n \sum_{p=1}^{P_n} g_{np} e^{\j \theta_{np}q}$.
Defining matrices $\vec{A}$ and $\vec{Z}$ elementwise as 
$[\vec{A}]_{mn}\defn a_{mn}$ and $[\vec{Z}]_{nq}\defn z_{nq}$, 
equation \eqref{Ybar2} reduces to $\ovec{Y}=\vec{AZ}+\ovec{W}$.

The above path-based parameterization of $\vec{Z}$ is not convenient because the angles $\{\theta_{np}\}_{p=1}^{P_n}$ vary over the users $n$ and are unknown.
Without loss of generality, we instead work with the critically sampled \cite{Sayeed:TSP:02} angles $\{2\pi l/\Nr\}_{l=0}^{\Nr-1}$, leading to 
\begin{align}
z_{nq}
&= \sum_{l=0}^{\Nr-1} x_{nl} e^{\j \frac{2\pi}{\Nr}lq} ,
\end{align}
where $x_{nl}$ can be interpreted as the $n$th user's contribution to the $l$th discrete receive direction.
Defining $\vec{X}$ elementwise as $[\vec{X}]_{nl}\defn x_{nl}$, we can write 
$\vec{Z}=\vec{XF}$ using DFT matrix $\vec{F}\in\Complex^{\Nr\times\Nr}$. 
Thus, after transforming the measurements $\ovec{Y}$ into the angle domain via $\vec{Y}\defn\ovec{Y}\vec{F}\herm/\Nr$ and $\vec{W}\defn\ovec{W}\vec{F}\herm/\Nr$, we obtain the linear model
\begin{align}
\vec{Y} = \vec{AX} + \vec{W} .
\label{eq:Y}
\end{align}

We note that, if each user contributed significantly to at most $D$ receive directions, then $\vec{X}$ would have at most $N=D\Nc\Nu$ significant coefficients, meaning that each column would have at most $D\Nc\Nu/\Nr$.
So, the columns of $\vec{X}$ become more sparse as the number of antennas $\Nr$ grows.

By restricting attention to a particular receive angle (or using a single-antenna BS), we obtain a model of the form 
$\vec{y}=\vec{Ax}+\vec{w}$, 
which coincides with the sparse linear inverse problem from \eqref{y}.

\bibliographystyle{ieeetr}
\bibliography{macros_abbrev,books,misc,comm,multicarrier,sparse,machine,stc,phase}

\end{document}